 \documentstyle[prb,aps,graphicx]{revtex}
\begin{document}
\draft

  \twocolumn
   \narrowtext

\title{ Spin dynamics in lightly doped La$_{2-x}$Sr$_x$CuO$_4$: \\
    Relaxation function within the $t - J$ model }
\author{Igor A. Larionov\cite{Auth1}}
\address{Magnetic Radiospectroscopy Laboratory, Department of Physics,
Kazan State University, 420008 Kazan, Russia}
\maketitle
\date{\today}

\begin{abstract}

The relaxation function theory of doped two-dimensional $S=1/2$ Heisenberg
antiferromagnetic (AF) system in the paramagnetic state is presented taking
into account the hole subsystem as well as both the electron and AF
correlations. The expression for fourth frequency moment of relaxation shape
function is derived within the $t-J$ model. The presentation obeys
rotational symmetry of the spin correlation functions and is valid for all
wave vectors through the Brillouin zone. The spin diffusion contribution to
relaxation rates is evaluated and is shown to play a significant role in
carrier free and doped antiferromagnet in agreement with exact
diagonalization calculations. At low temperatures the main contribution to
the nuclear spin-lattice relaxation rate, $^{63}(1/T_1)$, of plane $^{63}$Cu
arises from the AF fluctuations, and $^{17}(1/T_1)$, of plane $^{17}$O, has
the contributions from the wavevectors in the vicinity of $(\pi,\pi)$ and
small $q \sim 0$. It is shown that the theory is able to explain the main
features of experimental data on temperature and doping dependence of
$^{63}(1/T_1)$ in the paramagnetic state of both carrier free La$_2$CuO$_4$
and doped La$_{2-x}$Sr$_x$CuO$_4$ compounds.

\end{abstract}

\pacs{PACS numbers:  74.72.Dn, 74.25.Ha, 75.40.Gb, 71.27.+a }

\section{INTRODUCTION}

The spin dynamics in doped two-dimensional $S=1/2$ Heisenberg
antiferromagnetic (2DHAF) systems remains the one of the intriguing problems
of condensed matter in connection with physics of layered copper High
Temperature Superconductors (HTSC).\cite{Rigamonti_Progress} The effect of
doped holes in two-dimensional (2D) antiferromagnetic (AF) background was
studied in many papers.\cite{DagottoRMP} The Hohenberg-Mermin-Wagner theorem
states the absence of long range order in low-dimensional isotropic
Heisenberg systems at any finite temperature due to fluctuations, making the
order short-ranged. The temperature dependence of correlation length in
two-dimensional Heisenberg model was first described by Chakravarty,
Halperin and Nelson (CHN) by a quantum nonlinear $\sigma$ model\cite{CHN} in
accord with neutron scattering (NS) experiments\cite{NS_Exp_CHN_AZ} in
carrier free La$_2$CuO$_4$. A significant advance was achieved in
understanding the 2D Heisenberg systems at low temperatures due to the
improved further results of CHN in the renormalized classical
regime.\cite{HasenfratzNiedermayer,ChubSachYe} Since then the temperature
dependence of correlation length has been studied, e.g., by isotropic wave
theory\cite{Sokol_spin_wave} and by quantum consideration of
skyrmions.\cite{Kochelaev_Belov_xi}

Nuclear Quadrupole Resonance (NQR) and Nuclear Magnetic Resonance (NMR)
methods are very powerful in studying the low energy excitations of
HTSC.\cite{Rigamonti_Progress} Chakravarty and
Orbach\cite{ChakravartyOrbach} developed a theory of magnetic relaxation
phenomena in 2DHAF based on the quantum nonlinear $\sigma$ model of CHN in
the critical region of fluctuations. However, they considered only the
contribution that arises from the wave vectors in the vicinity of AF wave
vector $(\pi,\pi)$. Despite the fact that contribution from the wave vectors
${\bf q} \sim 0$ does not dominate in the plane copper spin-lattice
relaxation rate at low temperatures and was not accounted in the
analysis,\cite{ChakravartyBook1990} the contribution from spin diffusion
plays an important role in HTSC, especially in the plane oxygen relaxation
rate as measured by NMR, since the AF fluctuating contribution is filtered
out by the oxygen formfactor.

The Nearly Antiferromagnetic Fermi Liquid (NAFL) model of Millis, Monien and
Pines\cite{Pines_NAFL} reconciled the puzzling observation of non-Korringa
temperature dependence of the copper nuclear spin-lattice relaxation rate
and Korringa temperature dependence of the oxygen and yttrium nuclear
spin-lattice relaxation rates in optimally doped YBa$_2$Cu$_3$O$_7$ by
postulating both the localized Cu$^{2+}$ magnetic moments and free oxygen
holes. The NAFL model gave the relation between the AF correlation length
and the relaxation rates and was applied in a wide temperature and doping
range.\cite{Pines_Rev} However, this theory has some disadvantages,
connected mainly with the phenomenological character of the NAFL
description, since the temperature and doping dependence of correlation
length was postulated or, at best, taken from a comparison with experiment.

As a consequence, it is tempting to consider the 2DHAF systems doped by
charge carriers microscopically and the difficulties rise on. The
numerically exact methods study the relatively small clusters and in
addition, these methods are hardly applicable if we need to obtain the
dynamic quantities. In the absence of exact solution it is necessary to find
a reliable approach that will describe the physical quantities with
convincing accuracy. This is intriguing especially since the observation of
drastic change of various physical quantities with doping and emergence of
"stripe" physics in doped HTSC and related compounds.

The $t-J$ model became very popular since it was pointed by
Anderson\cite{Anderson_t_J} as a perspective model to describe the
electronic properties of layered HTSC cuprates. The present theory is
developed for paramagnetic state using a Mori-Zwanzig projection operator
procedure,\cite{Mori,Zwanzig} with a three-pole approximation for the
relaxation function.\cite{LoveseyMeserve} The merits of the theory were
demonstrated\cite{LoveseyMeserve} by comparison with experiments in
antiferromagnets in a wide temperature range down to temperatures close to
Neel temperature $T_N$. The advantage of the present formulation is that it
allows to take into account not only the AF correlation effects in
relaxation rates, but also the contribution from spin diffusion. The
presentation of the $t-J$ model in terms of Hubbard operators is known to
obey the rotational symmetry of the spin correlation functions and
automatically guarantees the exclusion of double occupancy. The form of
static susceptibility will be used from microscopic theory\cite{Zav98} as
obtained beyond the Random Phase Approximation (RPA) and naturally takes
into account the contribution from the hole subsystem. The dynamic structure
factor is a quantity directly measured in various experiments, e.g., by
magnetic neutron scattering. The advantages of expressing the dynamic
structure factor in terms of relaxation function were shown by Mori and
Kawasaki.\cite{MoriKawasaki} Until present the Mori-Zwanzig projection
operator procedure was applied in connection with HTSC only to carrier free
2D $S=1/2$ AF system\cite{chinaRel} with a three-pole approximation for the
relaxation function\cite{LoveseyMeserve} and the correlation length was used
from the results of CHN.\cite{CHN} It should be emphasized that the method
developed in the present work for calculations of dynamic structure factor
and nuclear spin-lattice relaxation rates is similar, however, the approach
is different from the calculations with the approximations for dynamic spin
susceptibilities.\cite{MoriyaPWA,Rigamonti_Progress,Pines_NAFL,Zav98,MyZavDB,Zav2001}

The paper is organized as follows. In Sec.~II, the basic relations are
presented for the relaxation function with a three pole approximation in a
continued fraction representation of the Laplace transform and the static
spin susceptibility. Sec.~III shows the evaluation of the second and the
fourth frequency moments of relaxation shape function within the $t-J$
model. Sec.~IV presents the results of calculations, comparison with
experiment, other theories and discussion. Sec.~V is the Conclusion.

\section{BASIC RELATIONS}

We employ the $t-J$ Hamiltonian written in terms of the Hubbard operators:
\begin{eqnarray} \label{Ht_J}
H_{t-J}=H_t + H_J & = & \sum_{i,j,\sigma}t_{ij}X_i^{\sigma 0}X_j^{0\sigma}
\cr & + &J\sum_{i>j}\left ({\bf S}_i{\bf S}_j-\frac{1}{4}n_i n_j\right ) .
\end{eqnarray}
Here, ${\bf S}_i$ are spin-1/2 operators at the lattice sites $i$, and
$X_i^{\sigma 0}$ are the Hubbard operators that create an electron with spin
$\sigma$ at site $i$.  The hopping integral $t_{ij}$ describes the motion of
electrons causing a change in their spins.  In this paper, $t_{ij}=t$ refers
to hopping between nearest neighbors and $J$ is the nearest-neighbor
antiferromagnetic (AF) coupling constant. The spin and density operators are
defined as follows:
\begin{equation} \label{SzSsigma}
S_i^{\sigma}=X_i^{\sigma \tilde \sigma}, \mbox{\hspace{6mm}}
S_i^z=\frac{1}{2}\sum_\sigma \sigma X_i^{\sigma \sigma},
\end{equation}
\begin{equation} \label{n_i_def}
n_i=\sum_\sigma X_i^{\sigma \sigma}, \mbox{\hspace{6mm}}
(\sigma =-\tilde \sigma ),
\end{equation}
with the standard normalization $X_i^{00}+X_i^{++}+X_i^{--}=1$. Without loss
of generality, we measure all energies from the "center of gravity" of the
band.

\subsection{Mori-Zwanzig projection operator procedure and a three pole
approximation for the dynamic relaxation function}

In the present formulation we will follow Mori.\cite{Mori} The time
evolution of a dynamical variable $S^z_{\bf k} (\tau )$, say, is given by
\begin{equation} \label{Liouville}
\dot{S}^z_{\bf k}(\tau ) \equiv \frac{d S^z_{\bf k}(\tau )}{d \tau} =
iLS^z_{\bf k}(\tau ).
\end{equation}
In general, $L$ is the Liouville operator, and in our, quantal case, $i L
S^z_{\bf k} (\tau )$ is the corresponding commutator with the Hamiltonian
(\ref{Ht_J}). The projection of the vector $ S^z_{\bf k} (\tau )$ onto the
$S^z_{\bf k}$ $\equiv$ $ S^z_{\bf k} (\tau =0)$ axis is given by
\begin{equation} \label{P0_def}
{\mathcal{P}}_0 S^z_{\bf k}(\tau ) = R({\bf k},\tau ) \cdot S^z_{\bf k},
\end{equation}
and defines the linear projection Hermitian operator ${\mathcal{P}}_0$. One
may separate $S^z_{\bf k} (\tau )$ into the projective and vertical
components with respect to the $S^z_{\bf k}$ axis:
\begin{equation} \label{S_projection}
S^z_{\bf k}(\tau ) = R({\bf k},\tau ) \cdot S^z_{\bf k} +
(1-{\mathcal{P}}_0)S^z_{\bf k}(\tau ),
\end{equation}
where
\begin{equation} \label{R_def}
R({\bf k},\tau ) \equiv (S^z_{\bf k}(\tau ),(S^z_{\bf -k})^{*}) \cdot
(S^z_{\bf k},(S^z_{\bf -k})^{*})^{-1},
\end{equation}
is the relaxation function in the inner-product bracket notation:
\begin{eqnarray} \label{inner_product}
 (S^z_{\bf k}(\tau ),(S^z_{\bf -k})^{*} & ) & \equiv  k_B T
\int_0^{1/k_B T} d \varrho
\cr & \times & \left< \exp(\varrho H)S^z_{\bf k}(\tau )
\exp(-\varrho H) (S^z_{\bf -k})^{*} \right>,
\end{eqnarray}
where the angular brackets denote the thermal average.

For future evaluations, it is convenient to introduce a set of quantities
$f_0(\tau )$, $f_1(\tau )$, $\dots$ , $f_j(\tau )$,
 $\dots \mbox{\hspace{1mm}}$ defined by equations
\begin{equation}
f_j(\tau ) \equiv \exp(i L_j \tau ) f_j \equiv \exp(i L_j \tau ) i L_j
f_{j-1},   \mbox{\hspace{2mm}} (j \geq 1),
\end{equation}
where  $f_0(\tau )$ $\equiv$ $S^z_{\bf k}(\tau )$, $L_j \equiv (1
-{\mathcal{P}}_{j-1}) L_{j-1}, \mbox{\hspace{1mm}} (L_0=L)$, and
\begin{equation}
\Delta_j^2 \equiv (f_j,f_j^{\mbox{\hspace{1mm}}*}) \cdot
(f_{j-1},f_{j-1}^{\mbox{\hspace{2mm}}*})^{-1}.
\end{equation}
The set $\{ f_j \}$ forms an orthogonal set. The larger number of $f_j$ is
used, the finer description of $S^z_{\bf k}(\tau )$ is obtained. The last
quantity from this set $f_n$, affected by evolution operator $\exp(i L_n
\tau )$, resulting in $f_n(\tau )$, was called the "n-th order random
force",\cite{Mori} acting on the variable $S^z_{\bf k}(\tau )$ and is
responsible for fluctuation from its average motion.

In terms of Laplace transform of the relaxation function, $R^L ({\bf k},\tau
)$, one may construct a continued fraction representation for $R ({\bf
k},s)$:
\begin{eqnarray} \label{R_L_3order}
R^L({\bf k},s) & = & \int_0^{\infty} d \tau \mbox{\hspace{1mm}} e^{s \tau} R
({\bf k},\tau )
\cr & = & 1/ \{ s+\Delta^2_{1{\bf k}}/[s+\Delta^2_{2{\bf k}}/(s+
\Delta^2_{3{\bf k}}/ \cdots )] \},
\end{eqnarray}
where $\Delta_{j {\bf k}}^2$ are related to the frequency moments
\begin{equation} \label{omega_nDef}
\left< \omega^n_{\bf k} \right> = \int_{-\infty}^{\infty}
d \omega \mbox{\hspace{1mm}} \omega^n F({\bf k},\omega) = \frac {1}{i^n}
\left[ \frac {d^n R({\bf k},\tau )}{d \tau^n} \right] _{\tau =0}
\end{equation}
of the relaxation shape function
\begin{eqnarray}
F( {\bf k }, \omega ) = \frac {1}{ \pi } & Re & [ R^L ({\bf k},i \omega) ]
\cr & = & \frac {1}{2\pi}  \int_{-\infty}^{\infty} d \tau \mbox{\hspace{1mm}}
e^{-i\omega \tau } R({\bf k},\tau ),
\end{eqnarray}
as
\begin{equation} \label{D1D2}
\Delta^2_{1{\bf k}}=\left< {\omega^2_{\bf k}} \right>, \mbox{\hspace{10mm}}
\Delta^2_{2{\bf k}}=\frac {\left< {\omega^4_{\bf k}} \right> }
{\left< {\omega^2_{\bf k}} \right> } - \left< {\omega^2_{\bf k}} \right>.
\end{equation}

 Lovesey and Meserve\cite{LoveseyMeserve} truncated the relaxation function
(\ref{R_L_3order}) to third order. They argued that since the ${\bf k}$
dependence of $\Delta_{2\bf k}$ is much weaker than that of $\Delta_{1\bf
k}$, and using the analytical results\cite{McFadden} for the sixth frequency
moment $\left< {\omega^6_{\bf k}} \right> $, the approximation of
$\Delta_{3\bf k}$ by a constant is a good approximation. Thus, they
suggested a three pole approximation for relaxation function,
\begin{equation} \label{R_3pole}
R({\bf k},s) = 1/\{s+\Delta^2_{1{\bf k}}/[s+\Delta^2_{2{\bf k}}/(s+1/
\tau_{\bf k})] \},
\end{equation}
with a cutoff characteristic time
\begin{equation} \label{tau_k}
\tau_{\bf k} = \left( \frac{2}{\pi \Delta^2_{2{\bf k}}} \right)^{1/2}.
\end{equation}
For $F({\bf k},\omega)$ this is equivalent to
\begin{equation} \label{FkwD1D2}
F({\bf k},\omega)=\frac{\tau_{\bf k} \Delta^2_{1{\bf k}}
\Delta^2_{2{\bf k}}/\pi} {[\omega \tau_{\bf k}
(\omega^2-\Delta^2_{1{\bf k}}-\Delta^2_{2{\bf k}})]^2 + (\omega^2 -
\Delta^2_{1{\bf k}})^2}.
\end{equation}
Here one should note that $F({\bf k},\omega)$ is real, normalized to unity
$\int_{-\infty}^{\infty} d \omega F({\bf k},\omega) = 1$ and even in both
${\bf k}$ and $\omega$.

The dynamic structure factor $S({\bf k},\omega)$ is related to the
relaxation shape function $F({\bf k},\omega)$ through the
fluctuation-dissipation theorem
\begin{equation} \label{SkwFkw}
S({\bf k},\omega)=\frac {2\pi \omega \chi({\bf k})}{1-\exp(-\omega / k_B T)}
F({\bf k},\omega).
\end{equation}

The only undefined quantity in the present formulation is the static spin
susceptibility $\chi({\bf k})$ in (\ref{SkwFkw}). Until present this method
was used\cite{LoveseyMeserve} to describe the paramagnetic state properties
of (anti)ferromagnets with the form of static susceptibility, which was
justified only at high temperatures (see also Refs.
\onlinecite{YoungShastry} and \onlinecite{chinaRel}).
In the present work we will employ the microscopic formula for static spin
susceptibility\cite{Zav98} that is shown to work in the overall temperature
range and properly takes into account the hole subsystem.

\subsection{Static spin susceptibility}

From Ref.~\onlinecite{Zav98}, the expression for static spin susceptibility
$\chi ({\bf k})$ is straightforward,
\begin{equation} \label{StatSusc}
\chi ({\bf k}) = \frac {4 | c_1 |}{J g_{-} (g_{+} + \gamma_{\bf k})},
\end{equation}
and its structure is the same as in the isotropic spin-wave
theory.\cite{Sokol_spin_wave} The meaning of $g_{+}$ is clear: it is related
to the correlation length $\xi$ via the expression
\begin{equation} \label{xi_gp}
\frac {\xi}{a} = \frac{1}{2 \sqrt{g_{+} -1}},
\end{equation}
where $a$ is a lattice unit. For carrier free AF system the relation
(\ref{xi_gp}) was obtained from the exponential decay of the spin-spin
correlation function at large separations, whereas at finite doping the same
expression was derived from the expansion of $\chi ({\bf k})$ taken around
the AF wave vector ${\bf Q}$, however, taking now into account the
contribution from the hole subsystem.\cite{Zav98}

\begin{equation} \label{c1c2def}
c_1=\frac{1}{z}\sum_\rho \langle S_i^zS_{i+\rho}^z\rangle ,
\mbox{\hspace{2mm}}
c_2=\frac{1}{z^2 - z}\sum_{\rho \neq \rho^\prime}\langle S_i^zS_{i+\rho
-\rho^\prime}^z\rangle ,
\end{equation}
are the nearest and next-nearest neighbor spin correlation functions,
respectively, the index $\rho$ runs over nearest neighbors, and
\begin{equation}
\gamma_{\bf k}=\frac{1}{z}\sum_{\bf \rho} \exp(i{{\bf k}  \rho})
=\frac{1}{2}(\cos k_x a +\cos k_y a),
\end{equation}
$g_{-} = 4\alpha z |c_1|$, $z$ = 4 is the number of nearest neighbors for
square lattice. The parameters $\alpha$ and $\beta$ were introduced in the
decoupling procedures for the higher-order Green's
functions.\cite{KondoYamaji} The parameter $\alpha$ preserves the important
property that spin operators obey the relation $\left<{\bf S}_i^2
\right>$=3/4 which should hold at all temperatures. The numerical values
for the temperature dependence of $\xi$ were determined in carrier free
La$_{2}$CuO$_4$ using $J = 0.12 $ eV and treating $\beta$ as the only
adjustable parameter.\cite{Zav98} The best fit to experimental data, which
were deduced from NS,\cite{NS_Exp_CHN_AZ} was obtained with $\beta = 2.5$.
This value will be kept fixed in the present calculations. The original
selfconsistent theory of Kondo and Yamaji (KY)\cite{KondoYamaji} with
$\alpha$ = $\beta$ = 1.705 fails in explanation of the absolute values of
$\xi$. Since $\beta$ enters in the combination $\beta c_2$, the increase of
the values of next-nearest correlations causes the extension of short-range
AF order and hence the enhancement of $\xi$ together with the spin stiffness
constant $\rho_S$.\cite{Zav98} In $T \rightarrow 0$ limit, for both the
carrier free and doped case, $c_2$ and $g_{-}$ are related as,
\begin{equation}
g_{-} = \frac {4}{3} (1+ 12 c_{2} \beta).
\end{equation}
 The reliability of the theory has been demonstrated by comparing the
numerical values for $c_1$, $c_2$ and $\chi_S \equiv \chi({\bf k}=0)$ with
Monte Carlo, Exact Diagonalization calculations and other
theories.\cite{Zav98}

In the present calculations for small doping $\delta$ we will use the
expression for doping and temperature dependence of $\xi$, given
by,\cite{Zav98}
\begin{equation} \label{xi_AZ}
\frac {\xi} {a} = \frac{J \sqrt{ \mbox{\hspace{1mm}} g_{-} } }{k_B T}
\exp (2\pi \rho_S/k_B T).
\end{equation}
As discussed in Ref.~\onlinecite{Zav98}, $\xi$ diverges at $T=0$ even in
doped samples and this disagrees with experiment. This disagreement appears,
probably, due to the overestimation of the role of AF correlations at low
temperatures in the KY decoupling procedure. To mimic the low temperature
behavior of the correlation length we will use the expression, as in
Refs.~\onlinecite{MyZavDB,Zav2001}, resulting in {\it effective} correlation
length $\xi_{\mathit eff}$, given by,
\begin{equation}\label{xi_eff}
\xi_{\mathit eff}^{-1}=\xi_0^{-1}+\xi^{-1}.
\end{equation}
Thus, the theory is able to explain the temperature and doping dependence of
correlation length. The expression (\ref{xi_eff}) is different from Keimer
{\it et al.}\cite{Keimer1992} empirical equation, where $\xi$ is given by
the Hasenfratz-Niedermayer formula\cite{HasenfratzNiedermayer} and hence,
there is no influence of the hole subsystem on $\xi$. In contrast, in the
present theory, $\xi$ is affected by doped holes. Thus from now on we
replace $\xi$ by $\xi_{\mathit eff}$.

\subsection{Excitation spectrum}

The band evolution with doping remains the controversial topic. The usual
parameter set in the $t-J$ model is $t=J/0.3$.\cite{DagottoRMP} The
electronic and AF spin-spin correlation functions reduce the
hoppings\cite{Plakida} resulting in {\it effective} values. In the early
proposal of $t-J$ model by Anderson\cite{Anderson_t_J,AndersonRVB_t_del} for
description of properties of layered copper HTSC compounds the
phenomenological relation was settled for the band width $\sim \delta t$.
Following the idea of Zhang and Rice\cite{ZhangRice} about copper-oxygen
singlets formation it was shown by Eremin {\it et al.},\cite{EreminBand}
that it is possible to describe correctly the elementary excitations
spectrum in cuprates. This singlet correlated band is analogous to upper
Hubbard band with essential distinction - the subband splitting is much
smaller compared to Hubbard model. Therefore it is possible to apply Hubbard
formalism without strict restriction on $t$ and $J$ values in $t-J$
model.\cite{EreminBand} Taking into account the AF spin-spin correlations
selfconsistently, resulting in {\it effective} hoppings, it was shown that
the band width varies linearly with doping in a wide range from lightly
doped to optimally doped compounds.\cite{MyZavDB} To avoid confusion from
approximations the simple expression for {\it effective} hopping
\begin{equation} \label{t_del_J_03}
t_{\mathit eff} = \delta J / 0.3,
\end{equation}
will be employed in the following evaluations to match the insulator - metal
transition. At high hole concentrations ($\delta \sim 1$), where the
correlation effects are negligible, Eq.~(\ref{t_del_J_03}) gives, as it
should, the value of band width for the noninteracting case.

Thus, $E_{\bf k}$ is given by
\begin{equation} \label{E_t}
E_{\bf k}=2t_{\mathit eff} (\cos{k_x a}+\cos{k_y a}).
\end{equation}
This expression resembles well also the values of singlet-correlated
bandwidth $\approx$ 0.4 eV (see Ref.~\onlinecite{EreminBand}) in optimally
doped ($\delta \approx 0.15$ per Cu site) layered copper HTSC as obtained
from Angle Resolved Photoemission Electron Spectroscopy
(ARPES).\cite{Gofron}

\section{Evaluation of expressions for frequency moments }

We now describe the procedure used to calculate the second $\left<
{\omega^2_{\bf k}} \right> $ and fourth $\left< {\omega^4_{\bf k}} \right> $
frequency moments of $F({\bf k},\omega)$, by calculating directly the
corresponding commutators in
\begin{equation} \label{omega2def}
\left< {\omega^2_{\bf k}} \right> = i \langle \left[ \dot{S}^z_{\bf k},
S^z_{-{\bf k}} \right] \rangle  / \chi({\bf k}),
\end{equation}
and
\begin{equation} \label{omega4def}
\left< {\omega^4_{\bf k}} \right> = i \langle \left[ \ddot{S}^z_{\bf k},
\dot{S}^z_{-{\bf k}} \right] \rangle  / \chi({\bf k}).
\end{equation}
The main effort necessary here is to obtain the expression for $\left<
{\omega^4_{\bf k}} \right> $. In this Section we will start with evaluation
of commutators and calculation of the thermodynamic averages, then introduce
the decoupling procedures and, finally, present the result for $\left<
{\omega^4_{\bf k}} \right> $. The procedure will be tested by comparison of
our expression for spin part with the existing result.\cite{LoveseyMeserve}
The $X_i^{0 \sigma}$ and $X_i^{\sigma 0}$ operators are fermions and obey
the anticommutation relations, whereas the $S_i^{\sigma}$ and $X_i^{\sigma
\sigma}$ are bosonic-like and obey the commutation relations. The terms
with the operators of different type are assumed to satisfy the commutation
relations. The commutators with the products of operators were decomposed on
terms that contain commutators and(or) anticommutators depending on the type
of operators.

\subsection{Evaluation of commutators}

In order to calculate the second $\left< {\omega^2_{\bf k}} \right>$ and the
fourth $\left< {\omega^4_{\bf k}} \right>$ frequency moments we first
evaluate the commutators
\begin{equation} \label{SzH_J}
\left[S^z_m,H_J \right]=\frac {1}{4} \sum_{j,\sigma} J_{mj} \sigma \left(
S_m^{\sigma}S_j^{\tilde{\sigma}}-S_j^{\sigma}S_m^{\tilde{\sigma}}\right),
\end{equation}
and
\begin{equation} \label{SzH_t}
\left[S^z_m,H_t \right]=\frac {1}{2} \sum_{j,\sigma} t_{mj} \sigma \left(
X_m^{\sigma 0} X_j^{0 \sigma} - X_j^{\sigma 0} X_m^{0 \sigma} \right).
\end{equation}

 The commutator of expression given by Eq.~(\ref{SzH_t}) with the hopping
term $H_t$ in (\ref{Ht_J}) is
\begin{eqnarray} \label{SzH_tH_t}
\mbox{\hspace{0mm}} & \mbox{\hspace{0mm}} & \left[
\left[ S^z_m,H_t \right ], H_t \right] =  \sum_{i,l,\sigma} t_{lm} \sigma
{\Big \{ } \frac {1}{2} t_{il} {\Big [ } X_m^{\sigma 0} X_l^{\tilde \sigma
\sigma} X_i^{0 \tilde \sigma} \cr & + & X_m^{\sigma 0}\left(
X_l^{00}+X_l^{\sigma \sigma}\right) X_i^{0 \sigma} + X_i^{\sigma 0}
\left( X_l^{00}+X_l^{\sigma \sigma}\right) X_m^{0 \sigma}
\cr & + & X_i^{\tilde \sigma 0} X_l^{\sigma \tilde \sigma}X_m^{0 \sigma}
{\Big ]} - t_{im} X_i^{\sigma 0}
\left( X_m^{00} + X_m^{\sigma \sigma} \right) X_l^{0 \sigma} {\Big \} },
\end{eqnarray}
whereas the commutator of expression given by Eq.~(\ref{SzH_J}) with the
spin part $H_J$ of (\ref{Ht_J}) is
\begin{eqnarray} \label{SzH_JH_J}
\mbox{\hspace{0mm}} & \mbox{\hspace{0mm}} &
\left[\left[ S^z_m,H_J \right ] , H_J \right] =
\frac {1}{8} \sum_{i,l,\sigma} J_{il} \left(J_{lm} - J_{im} \right)
\cr & \times & ( S_m^{\sigma} S_i^z S_l^{\tilde \sigma}
- S_m^{\sigma} S_i^{\tilde \sigma}
S_l^z + S_i^z S_l^{\sigma} S_m^{\tilde \sigma}
- S_i^{\sigma} S_l^z S_m^{\tilde \sigma} )\frac{}{}
\cr & + &  \frac {1}{8} \sum_{i,l,\sigma} J_{im} J_{lm}
(2S_i^{\sigma} S_m^z S_l^{\tilde \sigma}
- S_i^{\sigma} S_m^{\tilde \sigma} S_l^z
- S_i^z S_m^{\sigma} S_l^{\tilde \sigma}
\cr & + &  S_m^z S_l^{\sigma} S_i^{\tilde \sigma}
- S_m^{\sigma} S_l^z S_i^{\tilde \sigma}
+ S_l^{\sigma} S_i^{\tilde \sigma} S_m^z
- S_l^{\sigma} S_i^z S_m^{\tilde \sigma} ) .
\end{eqnarray}
The rest commutators of this type are as follows
\begin{eqnarray} \label{SzH_JH_t}
\cr &  & \left[\left[ S^z_m,H_J \right ], H_t \right] =
\frac{1}{4} \sum_{i,l,\sigma}  \sigma {\Big [} J_{im} t_{lm}
( X_m^{ \sigma 0} X_l^{0 \tilde \sigma } S_i^{ \tilde \sigma }
\cr &  & - S_i^{ \sigma } X_m^{ \tilde \sigma 0} X_l^{0 \sigma }
+ S_i^{ \sigma } X_l^{ \tilde \sigma 0} X_m^{0 \sigma }
-\frac{}{} X_l^{ \sigma 0} X_m^{0 \tilde \sigma } S_i^{\tilde \sigma })
\cr &  & + \left( J_{im} - J_{lm} \right) t_{il}
\left( S_m^{\sigma} X_i^{\tilde \sigma 0} X_l^{ 0 \sigma }-
X_i^{\sigma 0} X_l^{0 \tilde \sigma } S_m^{\tilde \sigma } \right) {\Big ]},
\end{eqnarray}
and
\begin{eqnarray} \label{SzH_tH_J}
\cr & & \left[\left[ S^z_m,H_t \right ], H_J \right] =
\frac{1}{8} \sum_{i,l,\sigma}  \sigma  t_{im} {\Big [ } J_{il}
( X_m^{ \sigma 0} X_i^{0 \tilde \sigma } S_l^{ \tilde \sigma }
\cr & & - X_m^{\sigma 0} X_i^{0 \sigma } X_l^{ \tilde \sigma \tilde \sigma}
+ X_i^{\tilde \sigma 0} S_l^{\sigma } X_m^{0 \sigma } \frac{}{}
- X_i^{ \sigma 0} X_l^{\tilde \sigma \tilde \sigma } X_m^{0 \sigma }
\cr & & + X_m^{\sigma 0} S_l^{\tilde \sigma } X_i^{0 \tilde \sigma }
- \frac{}{} X_m^{\sigma 0} X_l^{\tilde \sigma \tilde \sigma }X_i^{0 \sigma}
+ S_l^{ \sigma } X_i^{\tilde \sigma 0} X_m^{0 \sigma }
\cr & & - X_l^{\tilde \sigma \tilde \sigma } X_i^{\sigma 0} X_m^{0 \sigma})
- J_{lm}( X_m^{\tilde \sigma 0} S_l^{\sigma } X_i^{0 \sigma } \frac{}{}
- X_m^{ \sigma 0} X_l^{\tilde \sigma \tilde \sigma } X_i^{0 \sigma }
\cr & & + X_i^{ \sigma 0} X_m^{0 \tilde \sigma } S_l^{\tilde \sigma }
- \frac{}{} X_i^{\sigma 0} X_m^{0 \sigma} X_l^{\tilde \sigma \tilde \sigma}
+ S_l^{ \sigma } X_m^{\tilde \sigma 0 } X_i^{0 \sigma }
\cr & & - X_l^{\tilde \sigma \tilde \sigma } X_m^{\sigma 0} X_i^{0 \sigma} +
X_i^{ \sigma 0} S_l^{ \tilde \sigma } X_m^{0 \tilde \sigma } \frac{}{}-
X_i^{ \sigma 0} X_l^{\tilde \sigma \tilde \sigma } X_m^{0 \sigma }){\Big ]}.
\end{eqnarray}
We conclude this subsection with the emphasis, that our Eqs.
(\ref{SzH_t})-(\ref{SzH_tH_J}) are still exact. We will restrict further
ourselves and take into account the correlations between the first and the
second neighbors only. The resulting form of calculated commutators suggests
the types of necessary thermodynamic averages. The calculation of these
thermodynamic averages together with the decoupling procedures we need to
employ in order to estimate the values of higher spin, transfer amplitude
and density correlation functions will be presented in the following
subsections.

\subsection{Thermodynamic averages}

To calculate the thermodynamic averages, we use the retarded Green's
functions formalism. The equation of motion for a retarded Green's function
$\langle \langle A|B\rangle \rangle_\omega$ takes the form
\begin{equation} \label{GreenFuncDef}
\omega \langle \langle A|B\rangle \rangle_\omega=\langle [A,B]_
     +\rangle +\langle \langle [A,H]|B\rangle \rangle_\omega ,
\end{equation}
where $\left< ... \right> $ denotes the thermal average. The standard
relationship between correlation and Green's function may be written as
\begin{equation} \label{AverageGreen}
\langle BA\rangle =\frac{1}{2\pi i}\oint d\omega f(\omega )
                   \langle \langle A|B\rangle \rangle_\omega,
\end{equation}
where $f(\omega )=[\exp{(\omega /k_B T)}+1]^{-1}$ is the Fermi function; the
contour encircles the real axis without enclosing any poles of $f(\omega )$.

In general, Eq.~(\ref{GreenFuncDef}) cannot be solved exactly and one needs
some sort of approximation. To evaluate the Green's function $\langle
\langle [A,H]|B\rangle \rangle_\omega$ in Eq.~(\ref{GreenFuncDef}), one
uses a decoupling scheme originally proposed by Roth\cite{Roth} for
calculations on the Hubbard model. It can be shown that Roth's method is
essentially equivalent to the Mori-Zwanzig projection
technique\cite{Unger,Mehlig} and is strongly related to the moments method
as applied to the evaluation of the spectral density of the Green's
functions.\cite{Bowen,Harris} Roth's method has been studied by many
authors,\cite{Mehlig,Eremins1997,BeenenEdwards} and became a general method
to treat approximately the quasiparticle spectrum of an interacting system.
The reliability of the method has been demonstrated by comparison with the
exact diagonalization results.\cite{BeenenEdwards}

Roth's method\cite{Roth} implies that we seek a set of operators $A_n$,
which are believed to be the most relevant to describe the one-particle
excitations of the system of interest.  Also, it is assumed that, in some
approximation, these operators obey the relations\cite{Roth}
\begin{equation} \label{AKH}
[A_n,H]=\sum_mK_{nm}A_m,
\end{equation}
where the parameters $K_{nm}$ are derived through a set of
linear equations
\begin{equation} \label{K_calc}
\langle [[A_n,H],A_l]_+\rangle =\sum_mK_{nm}\langle
       [A_m,A_l^+]_+\rangle .
\end{equation}
Thus, it remains to define the operators $A_n$.  Because, in the framework
of the $t-J$ model, the quasiparticles are described by the Hubbard
operators $X_{\bf k}^{0\sigma}$, a set of operators $A_n$ contains only one
operator $A=X_{\bf k}^{0\sigma}$.  Hence, the matrix $K_{nm}$ is diagonal
and also contains one element $K=E_{\bf k}^\sigma$, where $E_{\bf k}^\sigma$
is the energy of of an electron with wave vector {\bf k} and spin projection
$\sigma$. Consequently, Eqs.~(\ref{AKH}) and (\ref{K_calc}) become
\begin{equation} \label{E_comm}
[X_{\bf k}^{0\sigma},H]=E_{\bf k}^\sigma X_{\bf k}^{0\sigma} ,
\end{equation}
\begin{equation} \label{E_AComm}
\langle [[X_{\bf k}^{0\sigma},H],X_{\bf k}^{\sigma 0}]_+\rangle =
E_{\bf k}^\sigma \langle [X_{\bf k}^{0\sigma},X_{\bf k}^{\sigma
0}]_+\rangle.
\end{equation}
In the 2D $t-J$ model, long-range order is absent at any finite temperature
and hence, $E_{\bf k}^\sigma$ does not depend on $\sigma$. Thus, we can
replace $E_{\bf k}^+$ and $E_{\bf k}^-$ by $E_{\bf k}$.

For our evaluations we need the thermal averages of the following types:
$\left< X_{i}^{\sigma 0} X_{j}^{0 \sigma} \right>$ and $\langle X_i^{\sigma
\sigma} X_{j}^{\sigma^\prime \sigma^\prime}\rangle$. The averages (spin-spin
correlation functions) of the type $\left< S_i^{\sigma } S_l^{\tilde \sigma}
\right>$ were defined in the SubSec.~II~B and the calculation procedure
together with the numerical values will be outlined in the Sec.~IV.

First, one should note, that in the absence of long-range order, $\langle
X_i^{\tilde \sigma \tilde \sigma}\rangle$ does not depend on the site index
and hence, according to Eq.~(\ref{n_i_def}), $T_0 = \langle X_i^{\tilde
\sigma \tilde \sigma}\rangle = \langle X_i^{\sigma \sigma}\rangle
=(1-\delta )/2$ and $c_0 = \langle S_r^z S_r^z \rangle = 1/4$.

The transfer amplitude between the first neighbors $T_1 = pI_1$ is given by
\begin{equation}
T_1 = pI_1 =- \frac{1}{z}\sum_{\rho} \left< X_{i}^{\sigma 0} X_{i+\rho}^{0
\sigma}
\right>
\end{equation}
and may be calculated using the spectral theorem
\begin{equation} \label{I_f_h}
I_1=-\sum_{\bf k}\frac{\gamma_{\bf k}}{e^{\frac{E_{\bf k}-\mu}{k_B T}}+1}
\equiv \sum_{\bf k} \gamma_{\bf k} f_{\bf k}^h.
\end{equation}
The parameter $I_1$ in Eq.~(\ref{I_f_h}) has been estimated in Ref.
\onlinecite{MyZavDB},
\begin{equation}
I_1 \approx \frac{4}{\pi}\left (1-e^{-\pi \tilde \delta}\right
)-2\tilde\delta,
  \mbox{\hspace{4mm}}\tilde \delta =\frac{\delta}{1+\delta},
\end{equation}
with an accuracy of a few percent over the whole region of $\delta$ from 0
to 1. Here one should note that for very small $\delta$ and low
temperatures, $I_1 \approx 2 \delta$.

The transfer amplitude between the second neighbors,
\begin{equation}
T_2 = \frac{1}{z(z-1)}\sum_{\rho \neq \rho \prime} \left< X_{i}^{\sigma 0}
X_{i+\rho-\rho \prime}^{0 \sigma} \right>,
\end{equation}
is
\begin{eqnarray}
\cr & T_2 = & \frac {p}{z(z-1)}\sum_{\bf k}\frac{16\gamma_{\bf k}^2 -
 4 \cos k_x a \cos k_y a -4}{e^{\frac{E_{\bf k}-\mu}{k_B T}}+1}
\cr & \equiv - & \frac{p}{z(z-1)} \sum_{\bf k} \left(16\gamma_{\bf k}^2 -
 4 \cos k_x a \cos k_y a -4 \right) f_{\bf k}^h.
\end{eqnarray}

For $p$ we have
\begin{equation}
p=\frac{1+\delta}{2},
\end{equation}
where $\delta$ is the number of {\em extra} holes, due to doping, per one
plane Cu$^{2+}$.  The chemical potential $\mu$ is related
to $\delta$ by
\begin{equation} \label{mudelta}
\delta = \frac{p}{N}\sum_{\bf k} f_{\bf k}^h.
\end{equation}
where $f_k^h=[\exp(-E_{\bf k}+\mu )/k_BT+1]^{-1}$ is the Fermi function of
holes.

To obtain the thermodynamic averages of the type $\langle X_i^{\sigma
\sigma} X_{i+\rho}^{\sigma^\prime \sigma^\prime}\rangle$ it is convenient to
make the following definitions
\begin{equation} \label{lambda_def}
\lambda = \lambda_{\tilde \sigma \tilde \sigma} =
\frac{1}{z}\sum_\rho \langle X_i^{\tilde \sigma \tilde \sigma}
X_{i+\rho}^{\tilde \sigma \tilde \sigma}\rangle,
\end{equation}
and
\begin{equation} \label{lambda_prime_def}
\lambda_{\sigma \sigma^\prime}=\frac{1}{z}\sum_\rho \langle
X_i^{\sigma \sigma} X_{i+\rho}^{\sigma^\prime \sigma^\prime}\rangle .
\end{equation}
To obtain $\lambda$ and $\lambda_{\sigma \tilde \sigma}$ we use the two
Green's functions\cite{MyZavDB}
\begin{equation}
G_{\bf k}^{(1)}(\omega )=\frac{1}{z}\sum_\rho \langle \langle X_{\bf
k}^{0\tilde \sigma}|X_i^{\tilde \sigma 0}X_{i+\rho}^{\tilde \sigma \tilde
\sigma} \rangle \rangle_\omega ,
\end{equation}
\begin{equation}
G_{\bf k}^{(2)}(\omega )=\frac{1}{z}\sum_\rho \langle \langle X_{\bf
k}^{0\tilde \sigma}|X_i^{\tilde \sigma 0}X_{i+\rho}^{\sigma \sigma}
\rangle \rangle_\omega .
\end{equation}
Note that in the paramagnetic state, $\lambda_{\sigma \sigma}=
\lambda_{\tilde \sigma \tilde \sigma}$ and
$\lambda_{\sigma \tilde \sigma}=\lambda_{\tilde \sigma \sigma}$.

According to Eqs. (\ref{GreenFuncDef}) and (\ref{E_comm}), the equation of
motion for $G_{\bf k}^{(1)}(\omega )$ and $G_{\bf k}^{(2)}(\omega )$ can be
written as
\begin{equation} \label{G1k_Eq_Motion}
(\omega -E_{\bf k})G_{\bf k}^{(1)}(\omega )=\frac{e^{i{\bf k}r_i}}{\sqrt{N}}
(1-p-\lambda_{\tilde \sigma \sigma}+pI\gamma_{\bf k})
\end{equation}
\begin{equation} \label{G2k_Eq_Motion}
(\omega -E_{\bf k})G_{\bf k}^{(2)}(\omega )=\frac{e^{i{\bf k}r_i}}{\sqrt{N}}
(1-p-\lambda_{\tilde \sigma \tilde \sigma}) .
\end{equation}
According to Eq.~(\ref{AverageGreen}):
\begin{equation}
 \lambda_{\tilde \sigma \tilde \sigma}
=\frac{1}{2\pi i}\sum_{\bf k} \frac{e^{-i{\bf k}r_i}}{\sqrt{N}}
\oint d\omega f(\omega )G_{\bf k}^{(1)}(\omega ),
\end{equation}
\begin{equation}
\lambda_{\tilde \sigma \sigma}
=\frac{1}{2\pi i}\sum_{\bf k} \frac{e^{-i{\bf k}r_i}}{\sqrt{N}}
\oint d\omega f(\omega )G_{\bf k}^{(2)}(\omega ) .
\end{equation}
Consequently, Eqs. (\ref{G1k_Eq_Motion}) and (\ref{G2k_Eq_Motion}) lead to a
system of linear equations for $\lambda_{\tilde \sigma \tilde \sigma}$ and
$\lambda_{\tilde \sigma \sigma}$ with the trivial solution
\begin{equation} \label{lambda_vs_p}
\lambda =\lambda_{\tilde \sigma \tilde \sigma}
=(1-p)^2-\frac{p^3}{2p-1}I^2,
\end{equation}
\begin{equation} \label{lambda_pm_vs_p}
\lambda_{\sigma \tilde \sigma}=(1-p-\lambda)
\frac{1-\delta}{1+\delta}=(1-p)^2+\frac{(1-p)p^2}{2p-1}I^2.
\end{equation}

\subsection{Decoupling procedures}

We now describe the decoupling procedures for the thermodynamic averages
performed in spirit of Hubbard and Jain\cite{HubbardJain} and Kondo and
Yamaji.\cite{KondoYamaji}

The averages of the type $\langle X_i^{\sigma 0}X_l^{0 \sigma} X_m^{\tilde
\sigma 0} X_j^{0 \tilde \sigma} \rangle $ are decoupled resulting in
products of transfer amplitudes and the decoupling parameter $\zeta$,
\begin{equation} \label{XsoXosXsoXos}
\langle X_i^{\sigma 0}X_l^{0 \sigma} X_m^{\tilde \sigma 0} X_j^{0 \tilde
\sigma} \rangle \rightarrow \zeta \langle X_i^{\sigma 0}X_l^{0 \sigma}
\rangle \langle X_m^{\tilde \sigma 0} X_j^{0 \tilde \sigma} \rangle.
\end{equation}

The four-spin correlation functions are approximated, as usually, by
products of two-spin correlation functions,\cite{LoveseyMeserve} however,
multiplied now with the decoupling parameter $\zeta$. Thus, we employ the
decoupling procedures
\begin{equation} \label{SSSSdecoupling}
\langle S_i^{\sigma }S_r^{\tilde \sigma} S_m^{\sigma } S_j^{ \tilde \sigma}
\rangle \rightarrow \zeta
\langle S_i^{\sigma }S_r^{\tilde \sigma} \rangle \langle S_m^{\sigma }
S_j^{ \tilde \sigma} \rangle,
\end{equation}
and
\begin{equation} \label{SzSzSSdecoupling}
\langle S_i^{z}S_r^{z} S_m^{\sigma } S_j^{ \tilde \sigma}\rangle
\rightarrow \zeta \langle S_i^{z}S_r^{z} \rangle \langle S_m^{\sigma }
S_j^{ \tilde \sigma} \rangle,
\end{equation}
for $i \neq r$ and $m \neq j$, whereas
\begin{equation} \label{c0SSdecoupling}
\langle S_r^{\sigma }S_r^{\tilde \sigma} S_m^{\sigma } S_j^{ \tilde \sigma}
\rangle \rightarrow 2c_0 \langle S_m^{\sigma } S_j^{ \tilde \sigma} \rangle.
\end{equation}

The averages with the products of operators $X_i^{\sigma 0}X_r^{0\sigma}$
between the nearest(next-nearest) neighbors with $(1-X_m^{\tilde \sigma
\tilde \sigma}) (1-X_j^{\sigma^\prime \sigma^\prime})$ are decoupled as
follows:
\begin{eqnarray}
\cr & \cr & \langle X_i^{\sigma 0}X_r^{0 \sigma} (1-X_m^{\tilde \sigma
\tilde \sigma}) (1-X_j^{\sigma^\prime \sigma^\prime} ) \rangle \nonumber \\
\cr & \cr & \rightarrow
\langle X_i^{\sigma 0}X_r^{0 \sigma} \rangle \langle 1 - X_m^{\tilde \sigma
\tilde \sigma} - X_j^{\sigma^\prime \sigma^\prime} + X_m^{\tilde \sigma
\tilde \sigma} X_j^{\sigma^\prime \sigma^\prime} \rangle. \label{XsoXos1_Xss}
\end{eqnarray}
and so on.

The averages with spin and Hubbard operators are decoupled as follows:
\begin{equation}\label{XsoXosSS}
\langle X_i^{\sigma 0}X_j^{0 \sigma} S_l^{\tilde \sigma} S_r^{\sigma} \rangle
\rightarrow \langle X_i^{\sigma 0}X_j^{0 \sigma} \rangle
\langle S_l^{\tilde \sigma} S_r^{\sigma} \rangle,
\end{equation}
and with spin and density operators:
\begin{equation} \label{XssSS}
\langle X_i^{\sigma \sigma} S_m^{\tilde \sigma} S_r^{\sigma} \rangle
\rightarrow \langle X_i^{\sigma \sigma} \rangle
\langle S_m^{\tilde \sigma} S_r^{\sigma} \rangle.
\end{equation}

The averages $\langle X_i^{\sigma \sigma} X_j^{\sigma^\prime \sigma^\prime}
\rangle$ between the second neighboring operators are decoupled simply by
\begin{equation} \label{Xss2decoupling}
\langle X_i^{\sigma \sigma} X_{i+2}^{\sigma^\prime \sigma^\prime} \rangle
\rightarrow \langle X_{i}^{\sigma \sigma} \rangle
\langle X_{i+2}^{\sigma^\prime \sigma^\prime} \rangle,
\end{equation}
because an inspection of Eqs.~(\ref{lambda_vs_p}) and (\ref{lambda_pm_vs_p})
shows that the values of averages of these type between the first neighbors
differ only slightly from $\left<X_i^{\sigma \sigma} \right>
\langle X_{i+\rho}^{\sigma \sigma} \rangle $. Therefore, the averages between
second neighbors in Eq.~(\ref{Xss2decoupling}) are thought as independent.
In addition, because the averages $\langle X_i^{\sigma \sigma}
X_j^{\sigma^\prime \sigma^\prime} \rangle$ between the first, in contrast to
next-nearest, neighbors, are calculated exactly, the averages like $\langle
X_r^{\sigma \sigma } X_m^{\sigma \sigma} X_j^{\sigma^\prime \sigma^\prime}
\rangle $ are decoupled in a way to avoid, where possible, the averages of
the type as given in Eq.~(\ref{Xss2decoupling}).

\subsection{Final result for $\left< {\omega^2_{\bf k}} \right> $ and
$\left< {\omega^4_{\bf k}} \right> $}

The expression for the second moment $\left< {\omega^2_{\bf k}} \right>$ is
straightforward. Calculating the commutator with the expressions given by
(\ref{SzH_J}) and (\ref{SzH_t}) with $S^z_r$, taking the thermal average and
the Fourier transform, the result is
\begin{equation} \label{w2}
\left< {\omega^2_{\bf k}} \right> =
- \left( 8 J c_1 - 4 t_{\mathit eff} p I_1 \right)
\left( 1- \gamma_{\bf k} \right) / \chi({\bf k}).
\end{equation}

We now proceed with calculating the commutators for $\left< {\omega^4_{\bf
k}} \right> $. Taking the commutators with the expressions given by Eqs.
(\ref{SzH_tH_t})-(\ref{SzH_tH_J}) with Eqs. (\ref{SzH_J}) and (\ref{SzH_t})
we obtain the expression for $\left< {\omega^4_{\bf k}} \right> $. The
expression was then approximated using the decoupling procedures for thermal
averages as was described in the previous SubSection. Taking the Fourier
transform we arrive to expression that contains various types of sums over
lattice sites. The corresponding types of sums over lattice sites are
presented in Appendix A. Utilizing these sums we obtain
\begin{eqnarray} \label{w4decoupled}
\cr & \cr & \left< {\omega^4_{\bf k}} \right> \simeq
-{\Big \{ } 128 J^3 [ c_2 \left(1-\gamma^2_{\bf k}\right) \left(\zeta
c_2\left(\gamma_{\bf k} - \frac{3}{4}\right) -\frac{1}{4} c_0 \right)
\cr & \cr & + c_0 c_1\left(\frac{7}{4} - \frac{5}{2}\gamma_{\bf k}
 + \frac{3}{4} \gamma^2_{\bf k} \right) + \zeta c_1 c_2 \left(\frac{13}{4} -
 \frac{15}{2}\gamma_{\bf k}+ \frac{17}{4}\gamma^2_{\bf k}\right)
\cr & \cr & + \zeta c^2_1\left(\frac{3}{2}-\frac{43}{8}\gamma_{\bf k}+\frac{21}{4}
\gamma^2_{\bf k}+ \frac{5}{8} \cos k_x a \cos k_y a- 2\gamma^3_{\bf k} \right)]
\cr & \cr & + 16 pI_1 t^3_{\mathit eff} [
c_1 \left(3 - 2 \gamma_{\bf k}^2 - \cos k_x a \cos k_y a \right)
\cr & \cr & +\zeta T_2 \left( 7 -12 \gamma_{\bf k} + 5 \gamma_{\bf k}^2 \right)
+ \frac {1-\delta}{2} \left(1-4\gamma_{\bf k} + 3 \gamma_{\bf k}^2 \right)
\cr & \cr & +  \left( \delta + \lambda \right) \left( - \frac {9}{2}
+ 9\gamma_{\bf k}-3\gamma_{\bf k}^2-\frac {3}{2}\cos k_x a \cos k_y a
\right)]
\cr & \cr & + 16 t_{\mathit eff} J^2 pI_1 [c_0 \left(- \frac {39}{8}
+ \frac{31}{4}\gamma_{\bf k}-\frac{23}{8}\gamma_{\bf k}^2 \right)
\cr & \cr & + c_1 \left( 16 \gamma_{\bf k}^3 - 35 \gamma_{\bf k}^2 +
25\gamma_{\bf k} - \frac{9}{2} - \frac{3}{2} \cos k_x a \cos k_y a \right)
\cr & \cr & + c_2 \left(-\frac {85}{8} +\frac {93}{4}\gamma_{\bf k}
- \frac {101}{8} \gamma_{\bf k}^2 \right)
+  \frac{9}{16} \frac{1-\delta}{2} \left( \gamma_{\bf k} - 1 \right)]
\cr & \cr & + 16 t^2_{\mathit eff} J [c_1 \left(\frac 34 \gamma_{\bf k}
-\frac 34 \right) + \frac {1+\delta}{2}  c_1
\left(6\gamma_{\bf k}^2 -\frac {45}{4} \gamma_{\bf k} +\frac {21}{4} \right)
\cr & \cr & +T_2 \left( \frac {1}{4}T_0 +\frac {3}{4} T_0^2 -\lambda \right)
\left(2\gamma_{\bf k}^2 -3\gamma_{\bf k} + 1 \right)
\cr & \cr & + \left( \frac {3}{4} \lambda^{+-} \left( 1- T_0 \right)
- T_0^2 +\lambda T_0 \right) \left( \gamma_{\bf k} - 1 \right)
\cr & \cr & + \frac {1+\delta}{2} c_2
\left(2 \gamma_{\bf k}^2 -\frac {9}{2} \gamma_{\bf k} +\frac {5}{2} \right)
 + c_1 T_0 \left(\frac {9}{4} \gamma_{\bf k}^2 -\frac {5}{2} \gamma_{\bf k}
+ \frac {1}{4} \right)
\cr & \cr & + c_1 T_2 \left(\frac {11}{4}\gamma_{\bf k}^2
-\frac {15}{2}\gamma_{\bf k} +\frac {19}{4} \right)
\cr & \cr & + c_2 T_0 \left(-2\gamma_{\bf k}^2 +\frac {9}{2}\gamma_{\bf k}-
\frac {5}{2}\right)
\cr & \cr & +\zeta (pI_1)^2 \left(-4 \gamma_{\bf k}^3 + 6 \gamma_{\bf k}^2 +
\frac {11}{4} \gamma_{\bf k} - \cos k_x a \cos k_y a - \frac {15}{4} \right)
\cr & \cr & + T_2 c_2 \left(16\gamma_{\bf k}^3 -21\gamma_{\bf k}^2
-\frac {5}{2}\gamma_{\bf k} +\frac {15}{2}\right)
\cr & \cr & + T_0 T_2 \left(2 \gamma_{\bf k}^2 -\frac {9}{2}\gamma_{\bf k}
+ \frac {5}{2} \right) + T_2 c_0 \left( -5 \gamma_{\bf k}^2
+ \frac {9}{2}\gamma_{\bf k} + \frac {1}{2} \right)
\cr & \cr & + \zeta T_2^2 \left(-2\gamma_{\bf k}^3 +6\gamma_{\bf k}^2
-\frac {19}{4}\gamma_{\bf k} +\frac {3}{4} \right)
 ] {\Big \} }  / \chi({\bf k}).
\end{eqnarray}

To insure the accuracy of the calculations we compare the result for spin
part $H_J$ with that reported in the literature.\cite{LoveseyMeserve} One
should note that $\left[S_m^z, n_i\right]=0$. Thus, the commutator
(\ref{SzH_J}) does not depend on the density-density term $n_i n_j$ in
(\ref{Ht_J}). Moreover, the commutator of spin operators $S_m^{\sigma}$
appearing in (\ref{SzH_J}) with the density operators $n_i$ is zero also:
$\left[S_m^{\sigma}, n_i \right] =0 $. This enables us to compare the result
(\ref{w4decoupled}) for spin part with the results for carrier free 2DHAF.
As it was already mentioned, we restrict ourselves and take into account the
correlations between the first and the next-nearest neighbors only.

For the second moment $\left< {\omega^2_{\bf k}} \right>$ the result
(\ref{w2}) is evidently correct. To compare the results for $\left<
{\omega^4_{\bf k}} \right>$ we need some additional effort. The
corresponding expressions and the procedure will be described in Appendix B.
The expression, as shown in (\ref{w4decoupled}), with $t_{\mathit eff}=0$
and with the decoupling parameter $\zeta$ settled to unity $\zeta = 1$ is
the same, as given in Ref.~\onlinecite{LoveseyMeserve} and may be compared
using the integrals (sums) over the Brillouin zone (see Appendix B).

\section{Comparison with experiment and discussion}

The results of the calculations are summarized in the Table~1. The value of
{\em extra} holes, due to doping, per one plane Cu$^{2+}$, $\delta$, can be
identified with the Sr content $x$ in La$_{2-x}$Sr$_{x}$CuO$_{4}$. The AF
spin-spin correlation functions $c_1$, $c_2$, the spin stiffness constant
$\rho_S$ and the parameter $g_{-}$ were calculated using the expressions and
the procedure as described in Ref.~\onlinecite{Zav98} (see also
Ref.~\onlinecite{MyZavDB}) in the $T \rightarrow 0$ limit, since we will
employ them in a temperature range $T$~$<$~1000~K~$\simeq$~0.7~$J$, where
experimental data exist and according to the calculations their values have
a weak temperature dependence.

\begin{table}

{\bf Table~1.} The calculated at $T\rightarrow 0$ values of AF spin-spin
correlation function $c_1$ between the first neighbors, the ${g_{-}}$
parameter and the spin stiffness constant $\rho_S$ as a function of
La$_{2-x}$Sr$_x$CuO$_4$ doping $x$ together with the values of decoupling
parameter $\zeta$ as extracted from comparison with $^{63}$Cu NQR
spin-lattice relaxation rate measurements.

$\phantom{\frac{\mbox{\hspace{1mm}}}{\mbox{\hspace{1mm}}}}$ {
\begin{tabular}
{l c c l c}
{\it x}
$\phantom{\frac{\mbox{\hspace{1mm}}}{\mbox{\hspace{1mm}}}}$ & $c_1$ & ${
g_{-} }$ & $2\pi \rho_S / J$ & $\zeta$ \\
\hline
0 & $-\phantom{\frac{\mbox{\hspace{1mm}}}{\mbox{\hspace{1mm}}}}$0.115215 &
4.1448 & 0.38 & 1.8  \\

0.012 & $-\phantom{\frac{\mbox{\hspace{1mm}}}{\mbox{\hspace{1mm}}}}$0.11474
& 4.137 & 0.37 & 1.8  \\

0.02 & $-\phantom{\frac{\mbox{\hspace{1mm}}}{\mbox{\hspace{1mm}}}}$0.11391 &
 4.117 & 0.365 &  -  \\

 0.024 & $-\phantom{\frac{\mbox{\hspace{1mm}}}{\mbox{\hspace{1mm}}}}$0.11333
& 4.102 & 0.36 & 1.6  \\

 0.03 & $-\phantom{\frac{\mbox{\hspace{1mm}}}{\mbox{\hspace{1mm}}}}$0.11238
&  4.080 &  0.355 & 1.5 \\

0.035 & $-\phantom{\frac{\mbox{\hspace{1mm}}}{\mbox{\hspace{1mm}}}}$0.11152
& 4.060 &  0.35 & $\sim$1.3 \\

 0.04 & $-\phantom{\frac{\mbox{\hspace{1mm}}}{\mbox{\hspace{1mm}}}}$0.11057
& 4.037 &  0.345 & 1.1 \\
\end{tabular}
}
\end{table}

\subsection{Temperature and doping dependence of antiferromagnetic
correlation length}

The AF correlation length, its doping and temperature dependence is given by
Eq.~(\ref{xi_eff}). $\xi_0$ is the value of correlation length at $T=0$ and
gives information on the topology of holes. At present, two types of doping
dependencies are under debate in the literature. The
first,\cite{Birgeneau_sqrt_x} is the localization of holes close to the
randomly distributed Sr$^{2+}$ ions gives $\xi_0 = a/\sqrt{x}$. The other
one is the formation of dynamic domain walls. In this case $\xi_0
= a / n x$, where $n$ is the average distance between the holes along the
domain walls in lattice units,\cite{Borsa_1_x,TranquadaStripe} usually
called "stripes".

\begin{figure}  [tbp]
    \centering \includegraphics[width=0.99\linewidth] {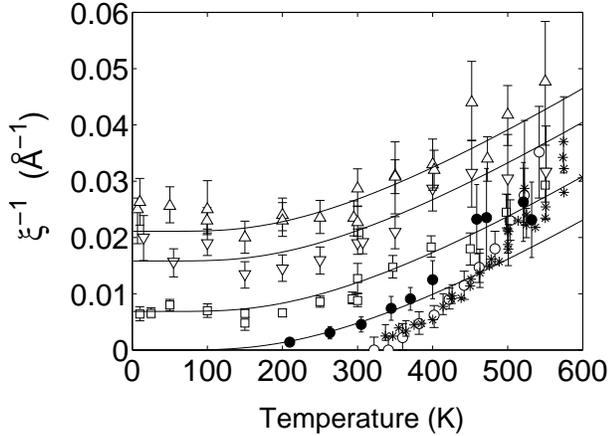}
\caption{The inverse correlation length $\xi_{\mathit eff}$ {\it versus}
temperature fitted (solid lines) to the experimental data as obtained from
neutron scattering experiments. For carrier free La$_{2}$CuO$_{4}$: filled
circles from Ref.~\protect\onlinecite{NS_Exp_CHN_AZ} (the fitted data),
asterisks from Ref.~\protect\onlinecite{MGrevenData} and open circles from
Ref.~\protect\onlinecite{Keimer1992}. For doped La$_{2-x}$Sr$_{x}$CuO$_{4}$:
squares for $x=0.02$, down triangles for $x=0.03$ and up triangles for
$x=0.04$ from Ref.~\protect\onlinecite{Keimer1992}.} \label{NS_Fig}
\end{figure}

The best fit of $\xi_{\mathit eff}$ to experimental data\cite{Keimer1992}
yields $\xi_0 = a / n x$, where $n=1.3$ for samples with $x \lesssim 0.02$
and $n=2$ for samples with $x > 0.02$, see Figure~\ref{NS_Fig}. These
results are in agreement with the the conclusion of Borsa {\it et
al.}\cite{Borsa_1_x} for La$_{2-x}$Sr$_{x}$CuO$_{4}$ compounds with $x
\lesssim 0.02$ and with the analysis of Carretta {\it et al.} on the basis
of dynamical scaling.\cite{CarrettaRigamontiSala} The presence of stripes
with $n=2$ was found also by Tranquada {\it et al.} in $x \simeq 1/8$
compound.\cite{TranquadaStripe}

  %%%   Figure 1

The fit with the relation $ a / n x$ in $\xi_0$ may be obtained by comparing
the experimental data in the low temperature region. Indeed, the doping
relation $a/\sqrt{x}$ fits with the neutron scattering data poorly than the
$ a / n x$ does, keeping in mind the value of plane lattice constant
$a=3.79$~\AA. For example, for $x=0.02$ one has $\xi_{0}^{-1} =
\sqrt{x} / 3.79 = 0.037$~\AA,$^{-1}$ whereas the experimental value is
0.005$-$0.007~\AA.$^{-1}$ The inability to fit the $1/\sqrt{x}$ relation
within the NS data\cite{Keimer1992} is clear also for larger values of
doping: $\xi_{0}^{-1}=\sqrt{x} / 3.79 = 0.046$~\AA$^{-1}$ for $x=0.03$ and
$\xi_{0}^{-1} = \sqrt{x} / 3.79 = 0.053$~\AA$^{-1}$ for $x=0.04$ differs
strongly from 0.013$-$0.02~\AA$^{-1}$ and 0.02$-$0.026~\AA,$^{-1}$,
consequently, as obtained by NS\cite{Keimer1992} (see Fig.~\ref{NS_Fig}). On
the other hand, the relation $\xi_{0}^{-1} = n x /a$ gives good agreement:
0.0069~\AA$^{-1}$ for $x=0.02$ with $n=1.3$, and 0.016~\AA$^{-1}$ for
$x=0.03$, and 0.021~\AA$^{-1}$ for $x=0.04$ with $n=2$. The value $n=1.3$
for $x=0.02$ fits the values of correlation length as obtained from
experimental data better in contrast to that with $n=2$. This result is
especially evident when one tries to compare the values of $\xi_{\mathit
eff}^{-1}$ at high temperatures. The high quality of the fit is in agreement
with previous studies and seems to confirm the microsegregation, however,
with different values of average distance between the holes along the domain
walls. It is tempting to speculate that the change in the average distance
$n$ from $n \approx 1.3$ for $x \lesssim x_c$ and $n=2$ for $x > x_c$
appears at the value of Sr content when the N\'eel order is completely
suppressed ($T_N \rightarrow 0$ at $x=x_c \simeq 0.02$) as discussed,
consequently, for example, in Refs.~\onlinecite{Borsa_1_x} and
\onlinecite{RigEurPhysJ1999}.

Having established the temperature and doping dependence of correlation
length we now proceed with calculations of the nuclear spin-lattice
relaxation rate.

\subsection{Spin-diffusion constant}

The spin-diffusion is described by the $({\bf k},\omega)=(0,0)$ mode. The
spin-diffusion plays an important role in an antiferromagnet and makes a
pronounced contribution to the relaxation rates. The spin-diffusion constant
$D$ is given by\cite{deGennesBennettMartin}
\begin{equation} \label{sDiffusion_def}
D = \lim_{{k} \rightarrow 0} \frac {1}{\pi {k}^2 F({\bf k},0)} =\lim_{k
\rightarrow 0} \frac {1}{k^2} \sqrt { \frac {\pi} {2} \frac {\left<
{\omega^2_{\bf k}} \right>^3 }{\left< {\omega^4_{\bf k}}\right>}}.
\end{equation}
For small ${\bf q}$, $\Delta_{1{\bf q}\rightarrow 0}^2$ tends to zero as
\begin{equation} \label{Del_1q0}
\Delta_{1{\bf q}\rightarrow 0}^2
= - (q_x^2 a^2 + q_y^2 a^2) (2Jc_1-t_{\mathit eff}pI_1) / \chi_S,
\end{equation}
whereas $\Delta_{2{\bf q}\rightarrow 0}^2 \simeq \left< \omega^4_{\bf q
\rightarrow 0} \right> /\left< \omega^2_{\bf q \rightarrow 0} \right> $
remains finite,
\begin{eqnarray} \label{Del_2q0}
\cr & \cr & \Delta_{2{\bf q}\rightarrow 0}^2
= {\Big \{ } 128J^3 {\Big [} \frac 18 c_2(\zeta c_2-c_0) -
\frac 14 \zeta c_1 c_2 + \frac 14 c_0 c_1
\cr & \cr &  - \frac{3}{32} \zeta c_1^2 {\Big ] } +
16pI_1t_{\mathit eff}^3 (\frac 32 c_1 + \frac 12 \zeta T_2 - \frac 12
\frac{1-\delta}{2})
\cr & \cr & + 16t_{\mathit eff}J^2pI_1(\frac 12 c_2
-\frac 12 c_0 -\frac{9}{64} \frac{1-\delta}{2} )
\cr & \cr & + 16t_{\mathit eff}^2 J {\Big [ }-\frac{3}{16}
\frac{1+\delta}{2} c_1 -\frac{1}{4}T_2 (\frac 14 T_0+\frac 34 T_0^2 -\lambda)
\cr & \cr & -\frac{1}{4}(\frac{3}{4}\lambda^{+-}(1-T_0)-T_0^2 +\lambda T_0 )
+ \frac{1}{8}c_2 \frac{1+\delta}{2} -\frac{3}{16}c_1
\cr & \cr & -\frac 12 c_1 T_0 -\frac 18 c_2 T_0
-\frac{3}{16} \zeta (pI_1)^2 +\frac 12 c_1 T_2 -\frac 78 T_2 c_2
\cr & \cr & +
\frac 18 T_2 T_0 +\frac{11}{8} T_2 c_0 -\frac {5}{16}\zeta T_2^2 {\Big ] }
{\Big \}} {\Big / } {\Big \{ } 2Jc_1-t_{\mathit eff}pI_1 {\Big \} }.
\end{eqnarray}

\begin{figure}  [tbp]
     \centering \includegraphics[width=0.99\linewidth] {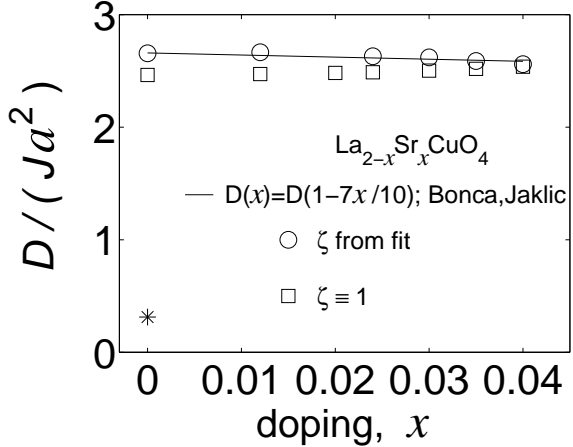}
\caption{The calculated doping dependence of the spin diffusion constant $D$
in the $T \rightarrow 0$ limit with the fixed decoupling parameter $\zeta=1$
(squares) and with $\zeta$ as obtained from the best fit to NQR data
(circles). The solid line is the result of Bonca and
Jaklic\cite{BoncaJaklic} for doping dependence of $D$ in the high
temperature limit. The asterisk marks the value of the spin diffusion
constant $D$ in the limit of high temperatures $T\rightarrow \infty$.}
\label{D_del_plot}
\end{figure}

The values of $D$ for carrier free AF system may be compared with the
results of other theories. The infinite temperature result is the same as in
Ref.~\onlinecite{chinaRel}, namely $D(T \rightarrow \infty) = \frac {1}{8}
\sqrt{2\pi} J a^2 = 0.3133 J a^2$. This value is close to that obtained by
Morita,\cite{Morita} $D=0.43 J a^2$ and $D= \frac {1}{10} \sqrt{5\pi} J a^2
=0.3963 J a^2$ in Refs.~\onlinecite{McFadden,Sokol_Spin_Diff,Collins}. The
calculated value of $D=2.46 J a^2$ at $T \rightarrow 0$ (with $\zeta=1$) is
larger compared to that obtained in Ref.~\onlinecite{chinaRel}, $D = 1.63 J
a^2$. The value of $D=2.66 J a^2$ (with $\zeta = 1.8$) is compatible with $D
\approx 3 J a^2$ obtained at $T=900$~K in Ref.~\onlinecite{Sokol_Spin_Diff}.
In general, $D$ weakly changes with doping and when one vary $\zeta$. Figure
\ref{D_del_plot} shows the doping dependence of $D$ for two cases, with
$\zeta=1$ and when $\zeta$ is obtained from the best fit to NQR data. The
calculated doping dependence of $D$ in the $T \rightarrow 0$ limit may be
compared with the results of Bonca and Jaklic\cite{BoncaJaklic} at high
temperatures, assuming that the doping dependence of $D$ remains the same.
Indeed, Figure~\ref{D_del_plot} shows the remarkable agreement.

   %%%   Figure 2

\subsection{Dynamic structure factor}

Now, since the nuclear spin-lattice relaxation rate is determined by the
dynamic structure factor $S({\bf q},\omega)$ we first consider its
temperature, frequency and wavevector dependence. First, we note that for
all temperatures the relaxation shape function $F({\bf q},\omega)$ as well
as $S({\bf q},\omega)$ give the elastic peak at $q=0$ and $\omega=0$. This
is clear from Eq.~(\ref{FkwD1D2}) since both $\left< {\omega^2_{\bf q}}
\right>$ and $\left< {\omega^4_{\bf q}} \right>$ $\sim q^2$ for small $q$.
We now turn to the case of finite but small NMR/NQR frequencies.

The relaxation shape function $F({\bf k},\omega)$ with small $\omega$,
compared to temperature scale of the system, related to $S({\bf k},\omega)$
as
\begin{equation} \label{Skw_Fkw_0}
S({\bf k},\omega \sim 0) = 2\pi k_B T \chi({\bf k}) F({\bf k},\omega),
\end{equation}
We now explore the form of $S({\bf k},\omega)$ and obtain two peaks: one at
${\bf q}\sim 0$ and the other at ${\bf Q} = (\pi /a ,\pi / a)$. The $S({\bf
q},\omega)$ value at small ${\bf q}$ is large when the ${\bf q}$ values are
such that
\begin{equation} \label{max_Fkw_D1D2_q0_def}
\Delta_{1{\bf q}}^2 \simeq \omega \tau_{{\bf q}} \Delta_{2{\bf q}}^2 ,
\end{equation}
see Eqs.~(\ref{FkwD1D2}),~(\ref{SkwFkw}), and
(\ref{Del_1q0}),~(\ref{Del_2q0}). Thus, $S({\bf q},\omega)$ has a sharp peak
at ${\bf q}_0$:
\begin{equation} \label{Fk0w_inv_w}
S({\bf q}_0 ,\omega) = \frac{k_B T \chi_S}{ \omega },
\end{equation}
and the $S({\bf k},\omega)$ value at ${\bf Q}=(\pi /a,\pi /a)$ is given by
\begin{equation} \label{FkQw_Del}
S({\bf Q},\omega) \simeq \frac{2 k_B T \chi({\bf Q})\tau_{{\bf Q}}
\Delta_{2{\bf Q} }^2 } { \Delta_{1{\bf Q} }^2 }.
\end{equation}
The relation given in Eq.~(\ref{Fk0w_inv_w}) is in agreement with the result
of Makivi\'{c} and Jarrel\cite{MakivicJarrel} on frequency dependence of the
dynamic structure factor at small values of wavevectors as extracted from
combination of the Maximum Entropy Method and Quantum Monte Carlo
calculations. This result agrees also with the basic relations known in the
literature. From general physical grounds, namely, linear response theory,
hydrodynamics and fluctuation-dissipation theorem, the diffusive spin
dynamics gives the form of dynamic structure factor\cite{ForsterKopietz}
\begin{equation} \label{S_Diff}
S({\bf q}\sim 0,\omega \sim 0) \simeq \frac {2 \chi_S} {1-\exp(-\omega
/k_BT)} \times \frac {\omega D {\bf q}^2}{\omega^2 +(D{\bf q}^2)^2},
\end{equation}
for small ${\bf q}$ and $\omega$.

Using Eq.~(\ref{S_Diff}) (or, equivalently, from
Eq.~(\ref{max_Fkw_D1D2_q0_def})) one may easy estimate the value of ${\bf
q}_0$ which is given by
\begin{equation} \label{q0_Diff}
{q}_0^2 \simeq \omega / D .
\end{equation}
For typical value of the measuring frequency, $\omega \approx$~1~mK, ${q}_0
a \approx \pi \times 10^{-4} $ and weakly changes when one vary doping and
the decoupling parameter $\zeta$ ($q_{0x}=q_{0y}=\frac{1}{\sqrt{2}} q_0 $).

For small ${\bf q} \ll {\bf q}_0 $ with finite $\omega$ the relaxation shape
function $F({\bf q},\omega)$ and the dynamic structure factor $S({\bf q},
\omega)$ approaches zero: $S({\bf q}_0 \gg {\bf q}\rightarrow 0,\omega)
\rightarrow 0$.

\begin{figure}  [tbp]
    \centering \includegraphics[width=0.99\linewidth] {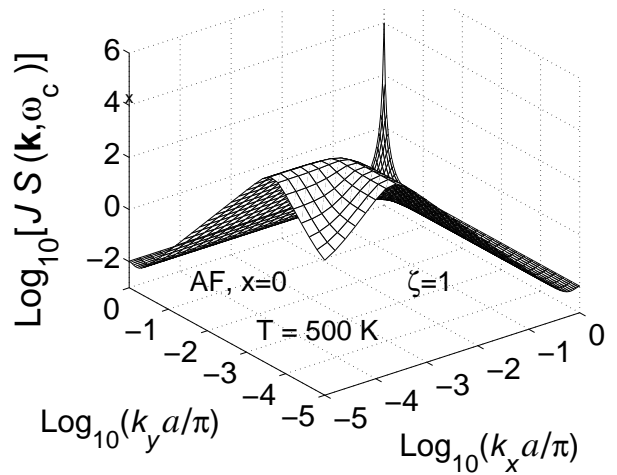}
\caption{Log-scale mesh of the calculated dynamic structure
factor $S({\bf k},\omega_c)$ for carrier free antiferromagnet at $T=500$~K
with simple decoupling $\zeta=1$. The cross on the vertical axis marks the
maximum of $J S({\bf k}, \omega_c) $ at ${\bf k} = {\bf q}_0$.}
\label{Skw_500Kx0AF_z1_Log3}
\end{figure}

\begin{figure}  [tbp]
     \centering \includegraphics[width=0.99\linewidth] {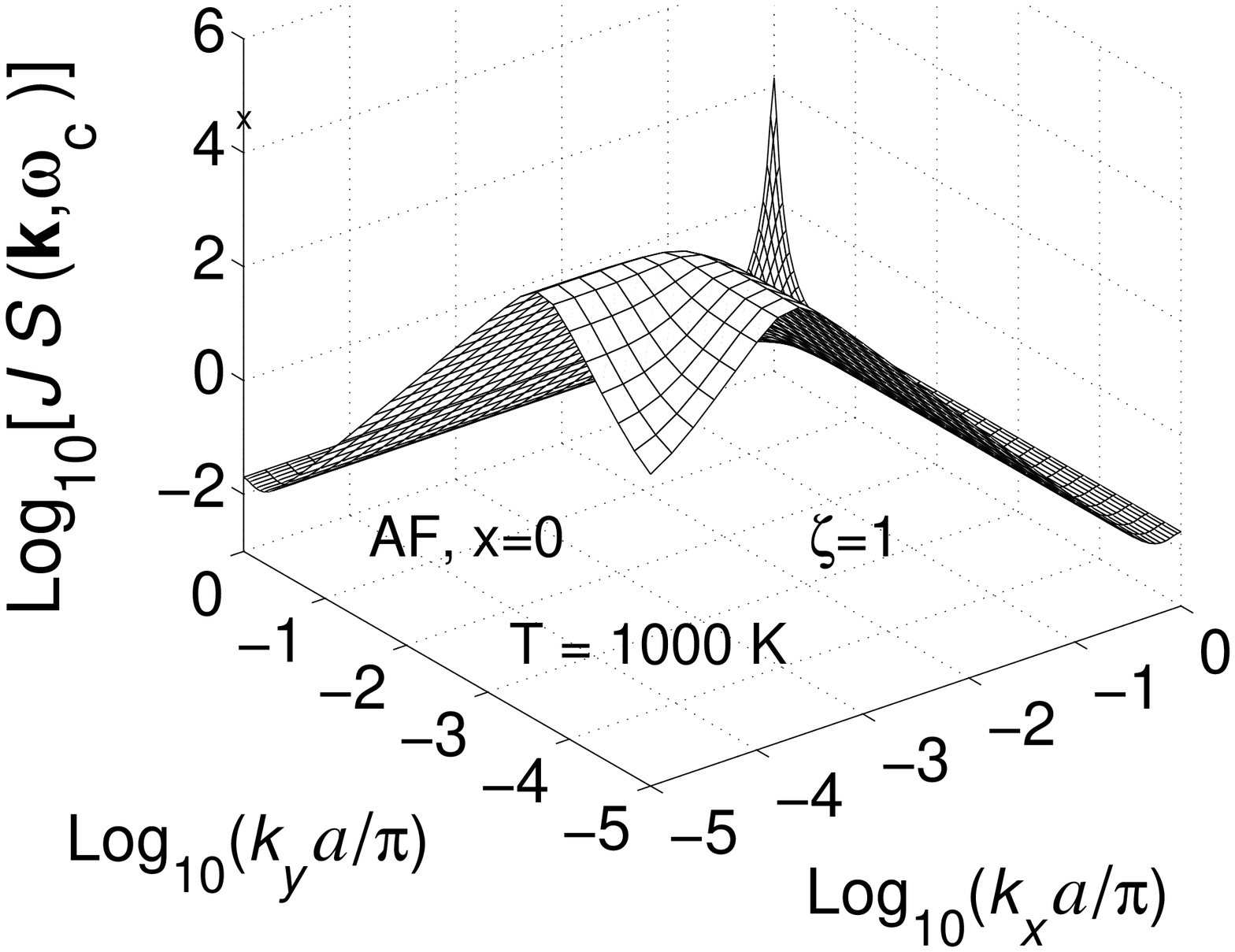}
\caption{Log-scale mesh of the calculated dynamic structure
factor $S({\bf k},\omega_c)$ for carrier free antiferromagnet at $T=1000$~K
with simple decoupling $\zeta=1$. The cross on the vertical axis marks the
maximum of $J S({\bf k}, \omega_c) $ at ${\bf k} = {\bf q}_0$.}
\label{Skw_1000Kx0AF_z1_Log3}
\end{figure}

\begin{figure}  [tbp]
     \centering \includegraphics[width=0.99\linewidth] {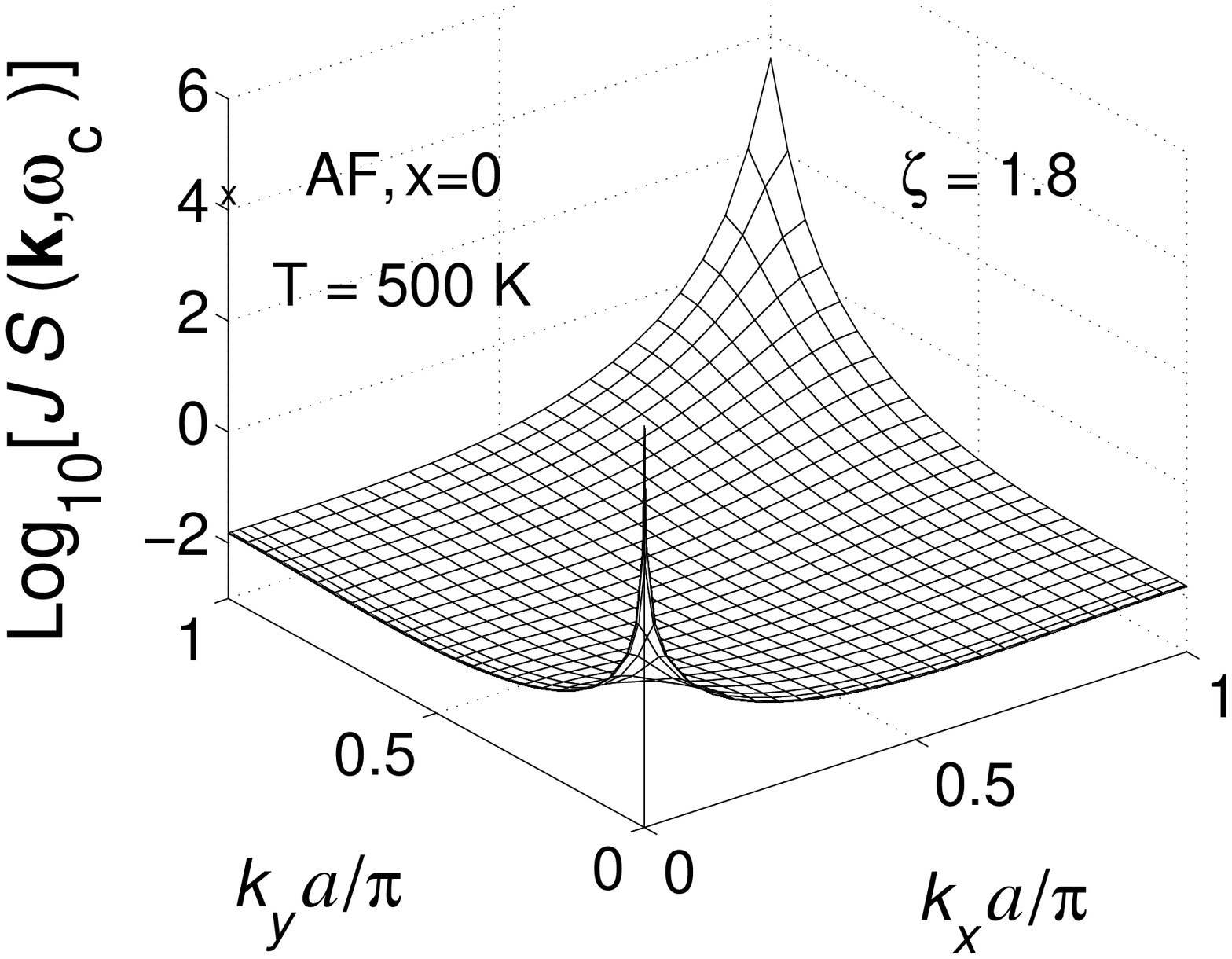}
\caption{Semilog-scale mesh of the calculated dynamic structure
factor $S({\bf k},\omega_c)$ for carrier free antiferromagnet at $T=500$~K
with the decoupling parameter $\zeta=1.8$. The region of small ${\bf q}$
values (${\bf q} \ll {\bf q}_0$) for which Log$_{10}[J S({\bf k},\omega)] <
-3$ is not shown. The cross on the vertical axis marks the maximum of $J
S({\bf k},\omega_c) $ at ${\bf k} = {\bf q}_0$.}
\label{Skw_500K_semiLog_AF_z18}
\end{figure}

\begin{figure}  [tbp]
      \centering \includegraphics[width=0.99\linewidth] {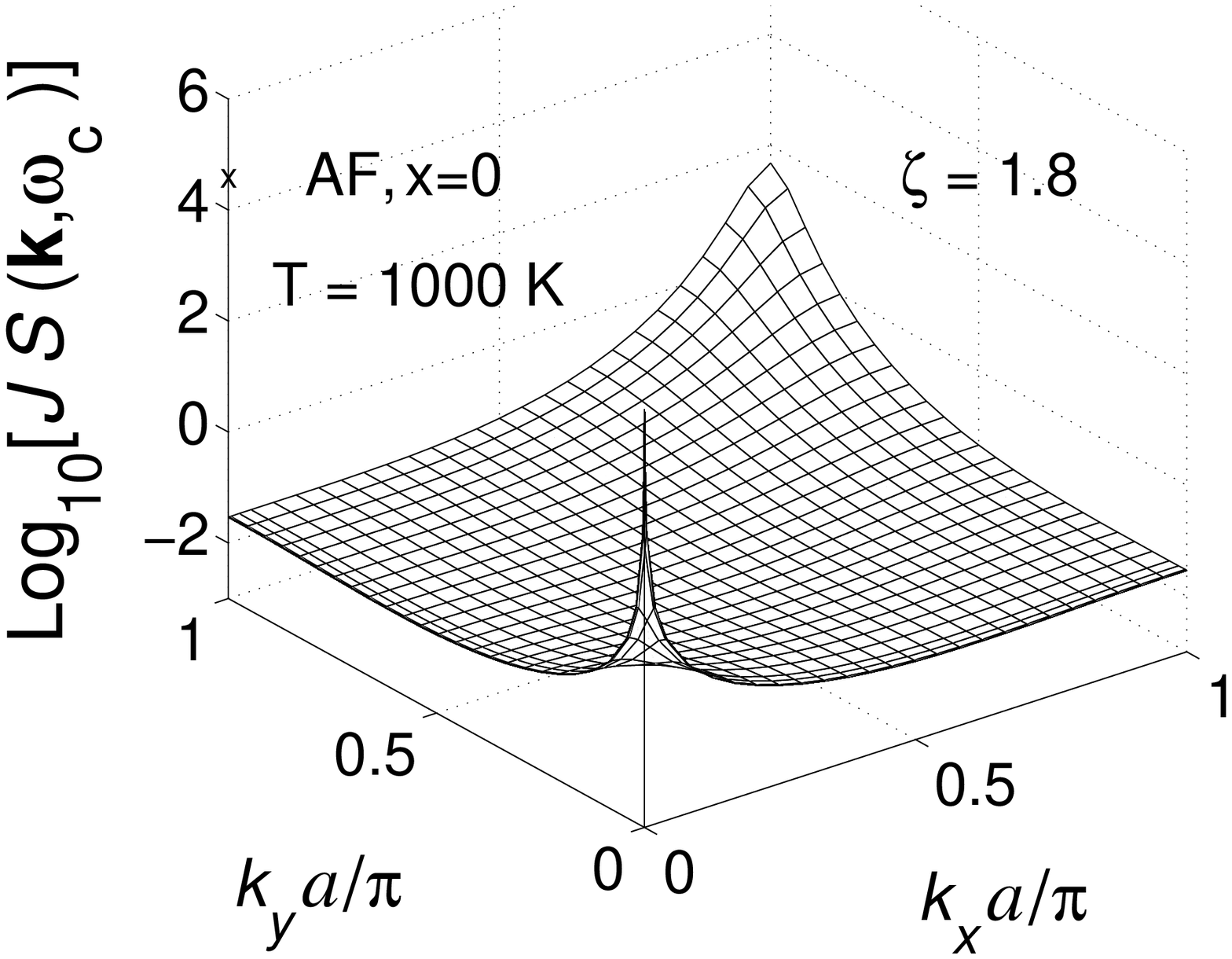}
\caption{Semilog-scale mesh of the calculated dynamic structure
factor $S({\bf k},\omega_c)$ for carrier free antiferromagnet at $T=1000$~K
with the decoupling parameter $\zeta=1.8$. The region of small ${\bf q}$
values (${\bf q} \ll {\bf q}_0$) for which Log$_{10}[J S({\bf k},\omega)] <
-3$ is not shown. The cross on the vertical axis marks the maximum of $J
S({\bf k},\omega_c) $ at ${\bf k} = {\bf q}_0$.}
\label{Skw_1000K_semiLog_AF_z18}
\end{figure}

\begin{figure}  [tbp]
     \centering \includegraphics[width=0.99\linewidth] {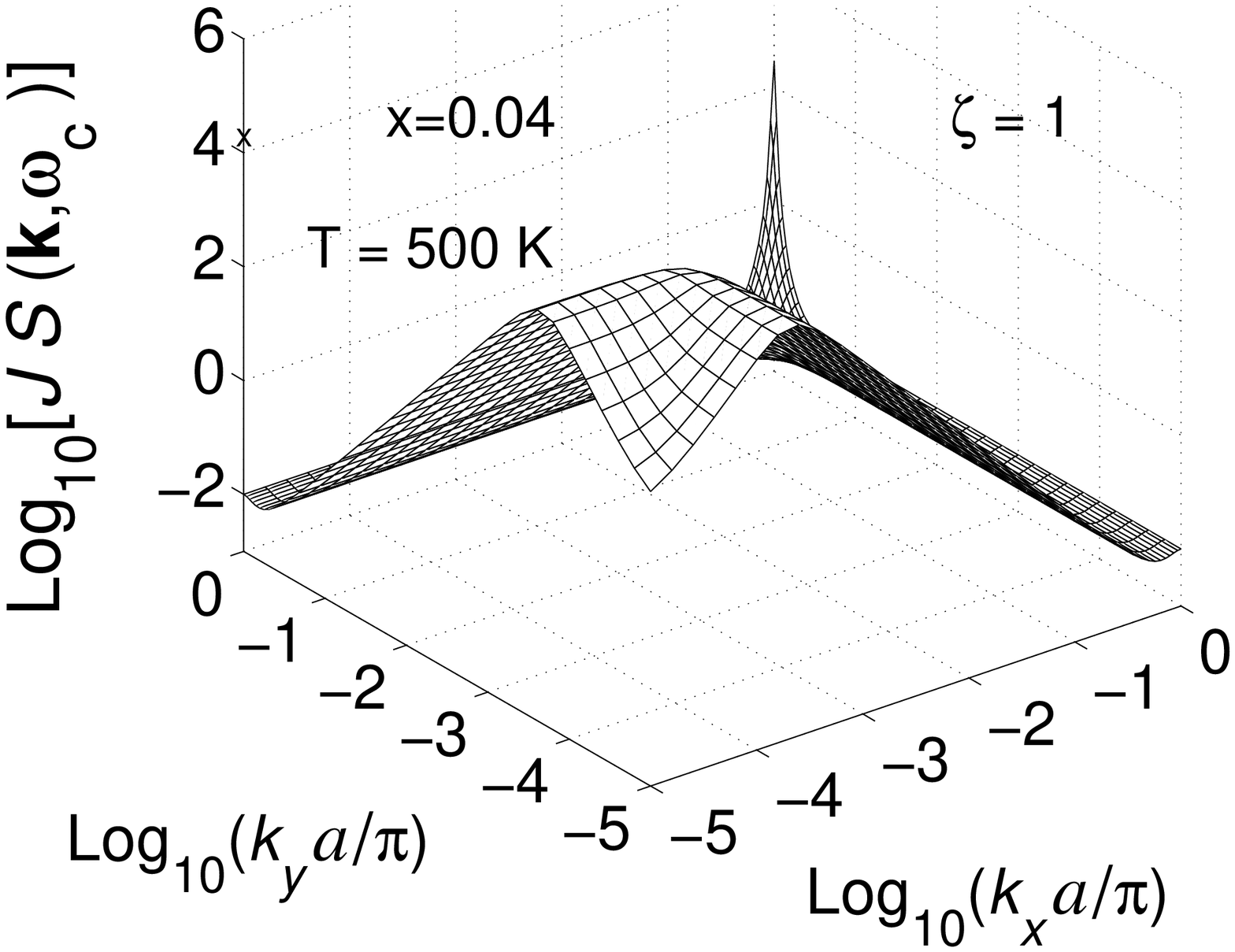}
\caption{Log-scale mesh of the calculated dynamic structure
factor $S({\bf k},\omega_c)$ for $x=0.04$ hole content at $T=500$~K with
simple decoupling $\zeta=1$. The cross on the vertical axis marks the
maximum of $J S({\bf k}, \omega_c) $ at ${\bf k} = {\bf q}_0$.}
\label{Skw_500K_Log3_x004_z1}
\end{figure}

\begin{figure}  [tbp]
     \centering \includegraphics[width=0.99\linewidth] {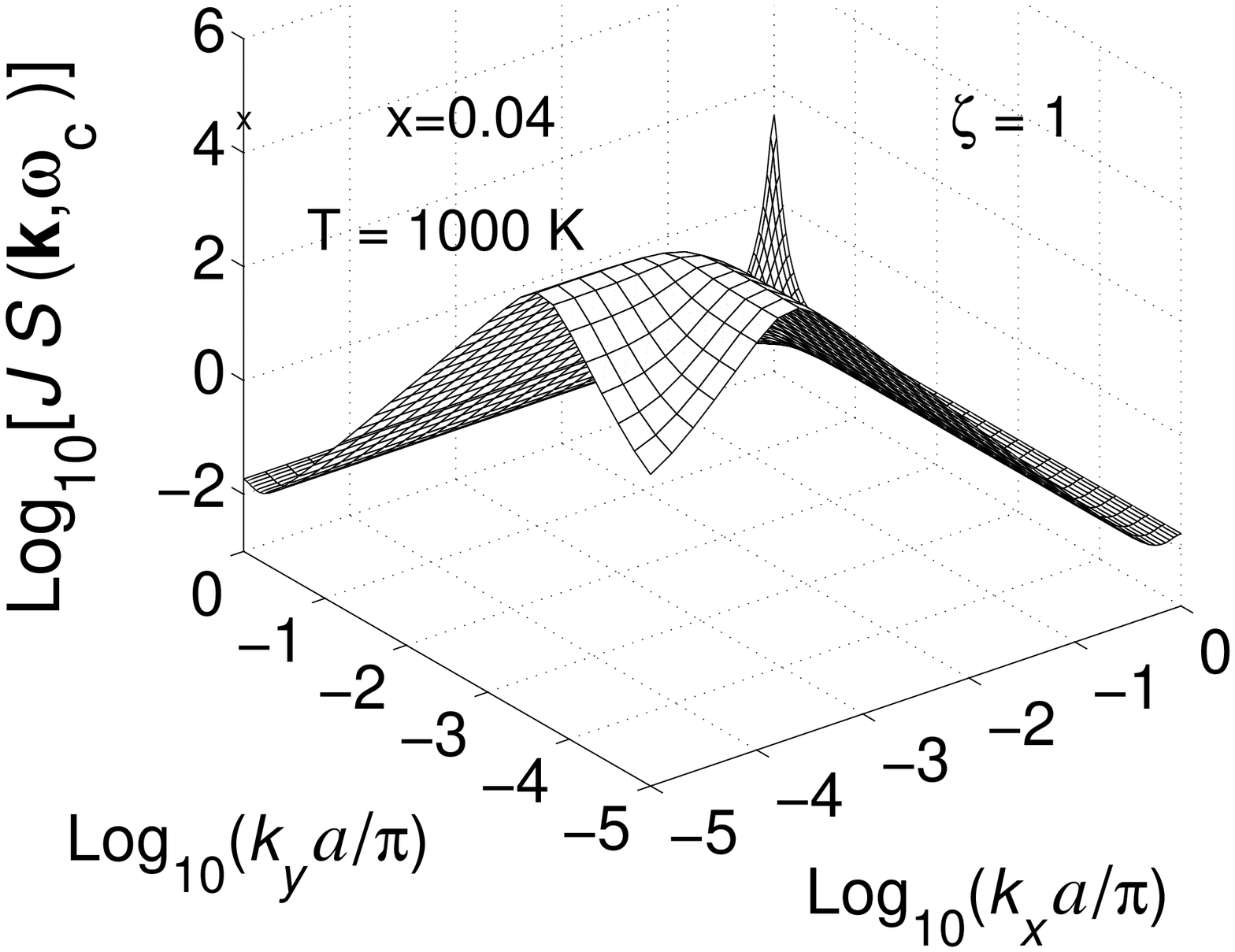}
\caption{Log-scale mesh of the calculated dynamic structure
factor $S({\bf k},\omega_c)$ for $x=0.04$ hole content at $T=1000$~K with
simple decoupling $\zeta=1$. The cross on the vertical axis marks the
maximum of $J S({\bf k}, \omega_c) $ at ${\bf k} = {\bf q}_0$.}
\label{Skw_1000K_Log3_x004_z1}
\end{figure}

\begin{figure}  [tbp]
     \centering \includegraphics[width=0.99\linewidth] {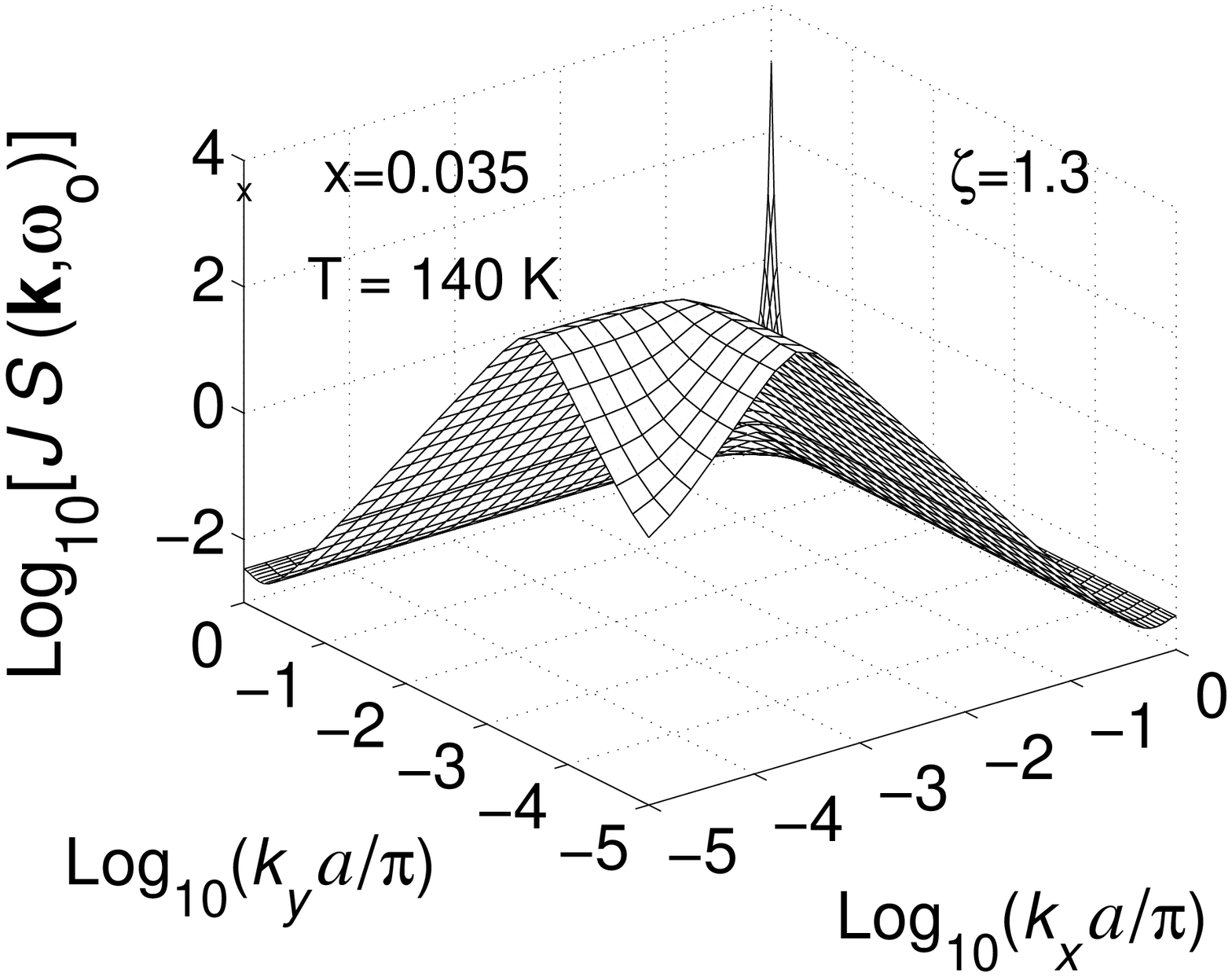}
\caption{Log-scale mesh of the calculated dynamic structure factor
$S({\bf k},\omega_0)$ for $x=0.035$ hole content at $T / J = 0.1$ ($T \simeq
140$~K) with simple decoupling $\zeta=1$. The cross on the vertical axis
marks the maximum of $J S({\bf k}, \omega_0) $ at ${\bf k} = {\bf q}_0$.}
\label{Skw_140K_Log3_x0035_z13}
\end{figure}

\begin{figure}  [tbp]
     \centering \includegraphics[width=0.99\linewidth] {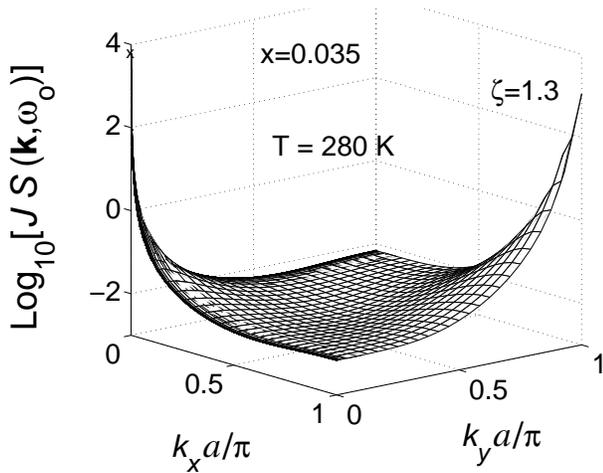}
\caption{Semilog-scale mesh of the calculated dynamic structure factor
$S({\bf k},\omega_0)$ for $x=0.035$ hole content at $T / J = 0.2$ ($T \simeq
280$~K). The region of small ${\bf q}$ values (${\bf q} \ll {\bf q}_0$) for
which Log$_{10}(J S({\bf k},\omega)) < -3$ is not shown. The cross on the
vertical axis marks the maximum of $J S({\bf k}, \omega_0) $ at ${\bf k} =
{\bf q}_0$. }
\label{Skw_280Kx0035_z13}
\end{figure}

Thus, the contribution to the nuclear relaxation rates from ${\bf q}$ around
$0$ has no peculiarities since $\omega$ is {\it finite}, but small, compared
to any values of the variables (we use $J$~=~0.12~eV~(1393~K)). Figures
(\ref{Skw_500Kx0AF_z1_Log3})-(\ref{Skw_280Kx0035_z13}) show the calculated
dynamic structure factor for different values of doping, temperature and
decoupling parameter $\zeta$ with $\omega_{\rm c} = 2\pi
\times$~33~MHz (~=~1.365~$\times$~10$^{-7}$~eV) and with $\omega_{\rm 0} =
2\pi \times$~52~MHz (~=~2.15~$\times$~10$^{-7}$~eV) for $x=0.035$.

    %%%   Figure 3

   %%%   Figure 4

     %%%   Figure 5

       %%%   Figure 6

    %%%   Figure 7

        %%%   Figure 8

         %%%   Figure 9

     %%%   Figure 10

\subsection{Nuclear spin-lattice relaxation}

The nuclear spin-lattice relaxation rate $^{\alpha}(1/T_1)$ is given by
\begin{equation} \label{T1ABSkw}
^{\alpha}( 1/T_1 ) = \frac{1}{\pi } \sum_{{\bf k}}
{ ^{\alpha} F ({\bf k})^2 } S({\bf k},\omega),
\end{equation}
where $\omega$ is the measuring NMR/NQR frequency, $^{\alpha}F({\bf k})$ is
the wave vector dependent hyperfine formfactor\cite{Mila,ShastryHypCoupl}
\begin{equation} \label{F_17O}
^{17} F ({\bf k})^2 = 2C^2 \left( 1 + \gamma_{\bf k} \right),
\end{equation}
for planar $^{17}$O and
\begin{equation} \label{F_63Cu}
^{63} F ({\bf k})^2 = \left(A_{ab} + 4\gamma_{\bf k} B \right)^2,
\end{equation}
for $^{63}$Cu sites. $A_{ab}$ and $B$ are the Cu on-site and transferred
hyperfine couplings, respectively. The quantization axis of the electric
field gradient coincides with the crystal axis $c$ which is perpendicular to
CuO$_2$ planes defined by $a$ and $b$. For $C$ in Eq.~(\ref{F_17O}), we
adopted the formula $C^2 = \frac{1}{6} (C_{\parallel}^2 + C_{\perp}^2)$,
where $C_{\parallel}$ and $C_{\perp}$ are the plane oxygen hyperfine
couplings for two axis perpendicular to $c$, and that includes the factor
$\frac{1}{3}$ to account $^{17}(1/T_1)$ as measured by NMR. The values of
hyperfine couplings were taken as follows:
$A_{ab}$~=~3.7$\times$10$^{-7}$~eV, for transferred hyperfine coupling $B$
the relation $B$~=~(1+2.75~$x$)$\times$3.8$\times$10$^{-7}$~eV is used to
match the weak changes with Sr doping\cite{Zha} and
$C$~=~2.8$\times$10$^{-7}$~eV in accord with the values as extracted from
NMR data and used in calculations of relaxation
rates.\cite{Rigamonti_Progress,Pines_NAFL,chinaRel,CarrettaRigamontiSala,RigEurPhysJ1999,Sokol_Spin_Diff,Mila,ShastryHypCoupl,Zha,MonienPines_C}

\begin{figure}  [tbp]
     \centering \includegraphics[width=0.99\linewidth] {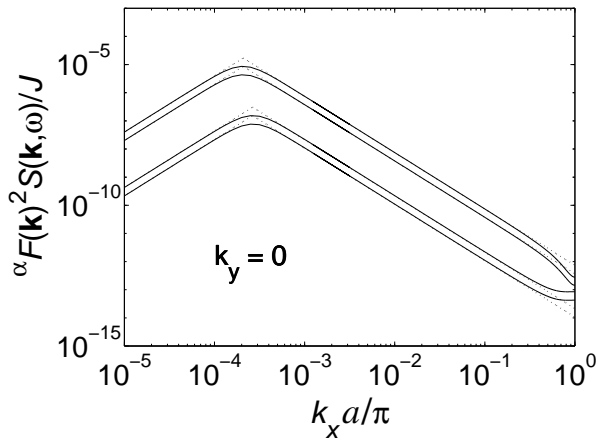}
\caption{Log-scale plot of the quantity $^{\alpha}F({\bf k})^2
S({\bf k},\omega) / J$ {\it versus} $k_x$ ($k_y = 0 $). The solid lines from
down till up: $^{17}F({\bf k})^2 S({\bf k},\omega_0) / J$ is given for
$x=0.035$ at $T/J=0.1$ ($T\simeq 140$~K) and $T/J=0.2$ ($T\simeq 280$~K)
with the decoupling parameter $\zeta=1.3$, $^{63}F({\bf k})^2 S({\bf
k},\omega_c) / J$ is given for carrier free AF ($x=0$) at $T=500$~K and
$T=1000$~K with the decoupling parameter $\zeta=1.8$. The dotted lines are
$^{\alpha}F({\bf q}_0)^2 S({\bf q}_0,\omega) {k_x}^2 / (0.5 J {\bf q}_0^2)$
for ${k_x} < {\bf q}_0$ and $^{\alpha} F({\bf q}_0)^2 S({\bf q}_0,\omega)
{\bf q}_0^2 / (0.5 J{k_x}^2)$ for $ k_x > {\bf q}_0$. }
\label{logFSkw_x_y0}
\end{figure}

\begin{figure}  [tbp]
     \centering \includegraphics[width=0.99\linewidth] {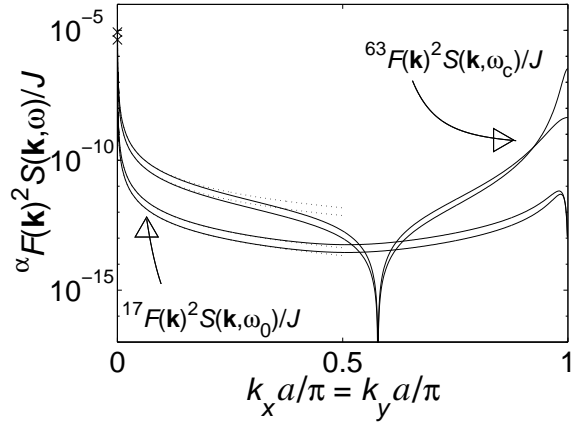}
\caption{Semilog-scale plot of the quantity $^{\alpha}F({\bf k})^2
S({\bf k},\omega) / J$ {\it versus} ${\bf k}$ along the diagonal of the
Brillouin zone ($k_x = k_y$). $^{17}F({\bf k})^2 S({\bf k},\omega_0) / J$ is
given for $x=0.035$ at $T/J=0.1$ ($T\simeq 140$~K, lower curve) and
$T/J=0.2$ ($T\simeq 280$~K, upper curve) with the decoupling parameter
$\zeta=1.3$. $^{63}F({\bf k})^2 S({\bf k},\omega_c) / J$ is given for
carrier free AF ($x=0$) at $T=500$~K (lower curve at small ${\bf k}$) and
$T=1000$~K (upper curve at small ${\bf k}$) with the decoupling parameter
$\zeta=1.8$. The dotted lines are $^{\alpha}F({\bf q}_0)^2 S({\bf
q}_0,\omega){\bf q}_0^2 / (0.5 J{\bf k}^2)$ for ${\bf k} > q_0$. The lower
and upper crosses mark the values of $^{63}F({\bf q}_0)^2 S({\bf
q}_0,\omega_c) / J$ at $T=500$~K and $T=1000$~K, respectively.}
\label{semilogFSkw_xy}
\end{figure}

The $^{17}$O and $^{63}$Cu nuclear relaxation rates are essentially
determined by the corresponding formfactors given by Eqs.~(\ref{F_17O}) and
(\ref{F_63Cu}). Figures~\ref{logFSkw_x_y0} and \ref{semilogFSkw_xy} show the
quantity $^{\alpha}F({\bf k})^2 S({\bf k},\omega)$ in a log-scale plot {\it
versus} $k_x$ ($k_y = 0$) and a semilog-scale plot along the diagonal of the
Brillouin zone. The calculations show the $1/{\bf k}^2$ wavevector
dependence of $S({\bf k},\omega)$ for ${\bf k} > {\bf q}_0$ and the ${\bf
k}^2$ dependence of $S({\bf k},\omega)$ for ${\bf k} < {\bf q}_0$. The form
of $^{\alpha} F({\bf k})^2 S({\bf k},\omega)$ gives the peaks at ${\bf q}_0$
and ${\bf k} \simeq (\pi /a,\pi /a)$. Thus, two types of contributions
dominate in the nuclear spin-lattice relaxation rates.

Here one should note, that the theory contains the decoupling procedures for
correlation functions and the decoupling parameter $\zeta$ has been
introduced. In general, despite the complicated structure of $\left<
\omega^4_{\bf k} \right>$, $\zeta$ allows to regulate the contributions with
${\bf k} \simeq (\pi /a, \pi /a)$ to the spin-lattice relaxation rate
$1/T_1$. The larger $\zeta$, the smaller is the contribution from ${\bf k}
\simeq (\pi /a, \pi /a)$ to $1/T_1$.

The contribution to $1/T_1$ from ${\bf q}$ around $0$ has no peculiarities
because the measuring frequency is {\it finite} ($\omega_{\rm c}
\approx$~$2\pi \times$~33~MHz (~=~1.58~mK) in $^{63}$Cu NQR and
$\omega_{\rm 0} \approx$~$2\pi \times$~52~MHz (~=~2.5~mK) in $^{17}$O NMR
measurements), but small, compared to any values of the variables (we take
$J$=0.12~eV~(1393~K)). The calculations also show that $^{\alpha} F({\bf
k})^2 S({\bf k},\omega)$ has a broad local minimum at ${\bf q}_m$ which
values are given by $q_{xm}^2 + q_{ym}^2 \approx \pi^2 / a^2$.

A direct numerical integration over ${\bf k}$ is difficult, because
$^{\alpha} F({\bf k})^2 S({\bf k},\omega)$ has an extremely sharp peak at
very small ${\bf q}_0$. This requires an unattainably large number of points
in numerical integration over Brillouin zone.

We first estimate the value of contribution to $1/T_1$ from small ${\bf q}$
because Figures~(\ref{logFSkw_x_y0}),~(\ref{semilogFSkw_xy}) show that the
pronounced contribution to relaxation rates should come from excitations
with $q \sim 0$. In the formulation of
Ref.~\onlinecite{ChakravartyBook1990}, the diffusive contribution,
\begin{equation} \label{FormulaChakravartyBook1990}
^{\rm Chakravarty} (1/T_1)_{\mathit Diff} \sim
\sum_{\bf q \sim 0}\frac{T\chi_S}{D{\bf q}^2}
\end{equation}
is divergent. While usually, this logarithmic divergence of
$(1/T_1)_{\mathit Diff}$ was argued to cut
off,\cite{ChakravartyBook1990,Sokol_Spin_Diff} since in a real systems the
mechanism destroying the diffusion should be taken into account. In
La$_{2-x}$Sr$_{x}$CuO$_{4}$ the combination of three-dimensional effects,
finite length scale and the presence of disorder in CuO$_2$ planes may
suppress the effect of spin diffusion on spin-lattice relaxation rates. For
example, using exact diagonalization technique in the high temperature
region, it was argued that strong local perturbation induced by oxygen
defects limit the spin-diffusion.\cite{Sokol_Spin_Diff} Despite of
considerable effort, the appropriate account of these effects is still
unavailable because of lack of any exact analytical result.

In the present theory the meaning of $q_0$ is clear: it is the value of
${\bf k}$ at which the maximum in $S({\bf k},\omega)$ occurs and for small
$\omega$ it may be treated as the cut off in the integration for calculation
of the contribution from spin diffusion (see
Eq.~(\ref{FormulaChakravartyBook1990}) and
Ref.~\onlinecite{ChakravartyBook1990}). For ${\bf q} \ll {\bf q}_0$ the
integration over ${\bf q}$ should be taken over the factor $\sim {\bf q}^2 /
{\bf q}_0^2 $. One should note, that in the present theory the form of the
calculated dynamic structure factor $S({\bf k},\omega)$ comes {\it not from,
e.g., disorder}, but rather than from the relaxation function and is in {\it
agreement} with linear response theory, hydrodynamics and
fluctuation-dissipation theorem and this result does not depend on the order
of the pole approximation in the relaxation function theory.

\begin{figure}  [tbp]
     \centering \includegraphics[width=0.99\linewidth] {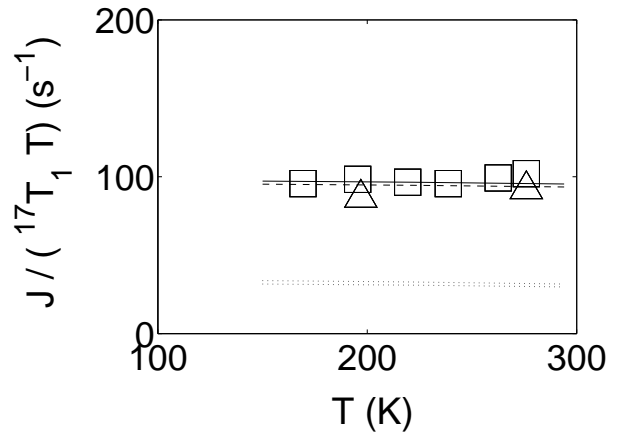}
\caption{The calculated temperature and doping dependence of the plane oxygen nuclear
spin-lattice relaxation rate $^{17}(1/T_1)$ (lines) and the experimental
data for La$_{2-x}$Sr$_{x}$CuO$_{4}$ as measured by NMR with $x=0.025$
(triangles) and $x=0.035$ (squares) from
Ref.~\protect\onlinecite{Imai1997_17O}. The experimental points have been
rearranged with $J = 1393$~K. The results of calculations with $\omega= 2
\pi \times$~52~MHz (9 Tesla) are given  for $x=0.035$ ($\zeta =1.3$) by
solid line and for $x=0.025$ ($\zeta =1.6$) by dashed line. The result of
calculation with $\omega= 2 \pi \times$~81.4~MHz for $x=0.035$ ($\zeta
=1.3$) coincides with the dashed line. The contribution to $^{17}(1/T_1)$
from the wave vectors in the vicinity of ($\pi/a,\pi/a$) for $x=0.035$ with
$\zeta=1$ is shown by upper dotted line and by lower dotted line with $\zeta
= 1.3$. }
\label{Rel_17O}
\end{figure}

Expanding $S({\bf q},\omega)$ around $q_0$ we obtain,
\begin{equation} \label{T1_inv_Diff}
^{\alpha} ( 1/T_1 )_{\mathit Diff}=\frac {^{\alpha} F(0)^2 k_B T a^2 \chi_S}
{\pi \hbar D}  { \Lambda },
\end{equation}
where $\Lambda$ depends on frequency through $q_0$. A simple and rough
estimate gives
\begin{equation} \label{Lambda_w}
{\large \Lambda } \sim \ln (1 / q_0^2)\sim \ln (const \times J/\omega).
\end{equation}
This result explains the reason why the oxygen $^{17}(1/T_1)$ relaxation
rate as measured by NMR remains unchanged at 9~T ($\omega_0 = 2 \pi
\times$~52~MHz) and 14.1~T ($\omega_0 ~= 2 \pi \times$~81.4~MHz) within the
experimental accuracy.\cite{Imai1997_17O} One should note that $\omega$ is
much less than $J$~=~1.8$\times$10$^8$~MHz, hence $\ln ( J / \rm 52~MHz) /
\ln (\it J / \rm 81.4~MHz) \approx$~1.03. Indeed, a sophisticated
calculation gives $\Lambda$(33~MHz)~=~2.52, $\Lambda$(52~MHz)~=~2.44, and
$\Lambda$(81.4~MHz)~=~2.37 and its value changes on less than 1\% when one
vary $\zeta$ and doping within $x \lesssim 0.04$. In view of the result that
the spin diffusion contribution is 70~\%, the relative shift of the measured
$^{17}(1/T_1)$ will be $\approx$~2~$\%$, that lies within the experimental
error (see Figure~\ref{Rel_17O}).

The $^{17}(1/T_1)$ relaxation rate has a contribution due antiferromagnetic
correlations between copper spins, however, for wave vectors only in the
vicinity of ${\bf Q} = (\pi /a,\pi /a)$ because the formfactor $^{17} F({\bf
k})$ filters out the contributions with ${\bf Q} = (\pi /a,\pi
/a)$. This filter causes also the minor sensitivity to the decoupling
parameter $\zeta$. The value of $\zeta = 1.3$ for $x=0.035$ (see Table~1) is
a plausible guess. This contribution was calculated by direct summation over
${\bf k}$ in the region ${\bf k} > {\bf q}_m$.

Figure \ref{Rel_17O} shows the calculated $^{17}$O relaxation rate in the
sector of lightly damped spin waves at low temperatures. It is seen that
$^{17}$O relaxation rate has a weak frequency dependence. The good agreement
of calculations (even without adjusting the parameter $\zeta$) with
experiment shows, that in case the mechanism destroying the diffusion is
present in La$_{2-x}$Sr$_{x}$CuO$_{4}$, whether it is caused by the
three-dimensional effects, finite length scale or the presence of disorder
in CuO$_2$ planes, it seems that it affects the spin-lattice relaxation
rates only little through the contribution from spin diffusion.

A fair agreement between experiment and calculated $^{17}(1/T_1)$ for
lightly doped La$_{2-x}$Sr$_{x}$CuO$_{4}$ may be thought as fortuitous
because of the following reason. In the present theory, the contribution to
relaxation rate from small ${\bf q}$ depends on spin diffusion constant $D$
as,
\begin{equation} \label{Rel_vs_D}
 ( 1/T_1 )_{\mathit Diff} \sim \frac {1}{D} {\ln(const \times D/\omega)} .
\end{equation}
In the $T\rightarrow 0$ limit the spin diffusion constant $D$ has to diverge
for both carrier free and lightly doped La$_{2-x}$Sr$_{x}$CuO$_{4}$ since
both Heisenberg and $t-J$ models have nonzero spin
stiffness.\cite{BoncaJaklic} On the other hand, in the quantum critical
region, $\rho_S < T < J $ (note that in the present theory $\rho_S \simeq
0.06 J \approx 80$~K and weakly decreases with light doping), the
spin-diffusion constant scales as $D \sim T^{1/2} \xi $ (see
Ref.~\onlinecite{CHN}). For lightly doped $x \approx 0.03$ systems one may
expect this scaling to be valid also, but with finite correlation length for
$T \gtrsim \rho_S $ (see Figure~\ref{NS_Fig}). This justifies the validity
of the present formulation and the propriety of the results obtained for
{\it finite} doping and low temperatures.

It is worth to mention the results of calculations by Chakravarty {\it et
al.}\cite{ChakravartyOrbach,CGKOW,KopietzChakravarty} of relaxation rates
for La$_2$CuO$_{4}$ on various nuclei. The results of CHN gave the
temperature dependence of correlation length with the two-loop corrections
in perfect agreement with experiment. The calculated plane copper relaxation
rate $^{63}(1/T_1)$ within the CHN theory is in agreement with
experiment,\cite{ImaiPRL1993} however, for plane oxygen the results of
calculations does not reproduce the values of the measured $^{17}(1/T_1)$
for insulating Sr$_2$CuO$_2$Cl$_2$ (see Ref.~\onlinecite{Imai1997_17O}).
Moreover, the agreement between theory and experiment becomes worse when the
two-loop correction has been taken into account.\cite{KopietzChakravarty} A
possible reason of this inconsistency is the missed contribution from spin
diffusion\cite{KopietzChakravarty} and the electronic structure of
Sr$_2$CuO$_2$Cl$_2$ with pockets centered around $(\pi/2,\pi/2)$ as observed
by LaRosa {\it et al.}\cite{LaRosa} Thus, for Sr$_2$CuO$_2$Cl$_2$ one has to
take into account the hoppings between the next nearest and next-next
nearest neighbors that are beyond our present consideration.

\begin{figure}  [tbp]
     \centering \includegraphics[width=0.99\linewidth] {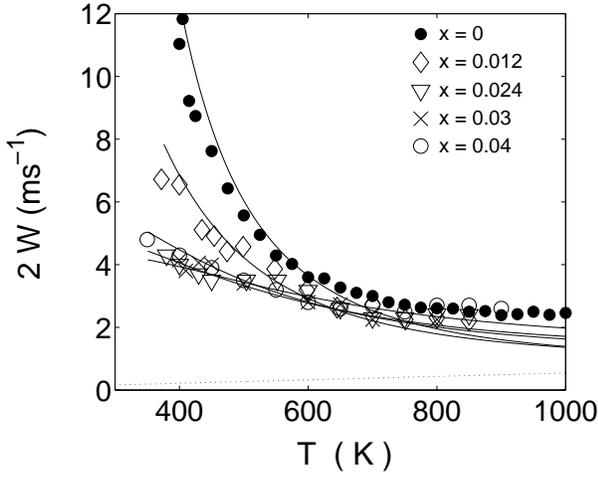}
\caption{Temperature and doping dependence of the plane copper nuclear
spin-lattice relaxation rate $^{63}(1/T_1) = 2W$ (solid lines) fitted to the
experimental data for La$_{2-x}$Sr$_{x}$CuO$_{4}$ with $x=0$, $x=0.012$,
$x=0.024$ and $x=0.03$ from Ref.~\protect\onlinecite{RigEurPhysJ1999} and
with $x=0.04$ from Ref.~\protect\onlinecite{ImaiPRL1993}. The values of the
fitting parameter $\zeta$ are shown in the Table~1. The dashed line shows
the contribution to $^{63}(1/T_1)$ from spin diffusion for $x=0$. }
\label{Rel_63Cu}
\end{figure}

We now discuss the temperature and doping dependence of plane copper nuclear
spin-lattice relaxation rate $^{63}(1/T_1)$. Figure \ref{Rel_63Cu} shows the
fitted $^{63}(1/T_1)$ to experimental data in lightly doped
La$_{2-x}$Sr$_{x}$CuO$_{4}$. It is seen that the agreement is good for
temperatures $T \lesssim J/2 $. For temperatures $ T > J/2$ the agreement is
less satisfactory. This is a bit puzzling but could be due to better account
of the $\xi$ values in the low temperature region ($T < J/2$) compared to $T
\approx J$ and the preexponential factor $\sim 1/T$ in (\ref{xi_AZ}) and in
the spin-wave theory,\cite{Sokol_spin_wave} which is an artifact of the
mean-field approach. For carrier free La$_{2}$CuO$_{4}$ and at $T=1000$~K
($T \simeq 0.7 J $), $\xi \approx $~5$a$. Obviously, the value of
correlation length $\xi$, as extracted in the limit of large
separations\cite{Zav98} satisfies the inequality $\xi^2 /a^2 \gg 1$. At
temperatures $T < J$ the behavior of $^{63}(1/T_1)$ is smooth: the slope of
the curve decreases with increasing $T$. At high temperatures,
$^{63}(1/T_1)$ is expected to have a minimum at $T \approx J$ and to be
dominated by spin diffusion at $T \geq J $. The validity of
Eq.~(\ref{xi_AZ}) does not allow to calculate $1/T_1$ at $T
> J$. Similar temperature dependence of $^{63}(1/T_1)$ was found in
Ref.~\onlinecite{chinaRel}. It was found that the saturation of
$^{63}(1/T_1)$ at $T\sim J/2$ is "primarily due to a competition from the
spin diffusion over the critical slowing down". The present work shows that
the overwhelming contribution arises from strong short-range
antiferromagnetic correlations between copper spins and at low temperatures
the contribution from spin diffusion to $^{63}(1/T_1)$ is small (see
Figure~\ref{Rel_63Cu}).

\begin{figure}  [tbp]
     \centering \includegraphics[width=0.99\linewidth] {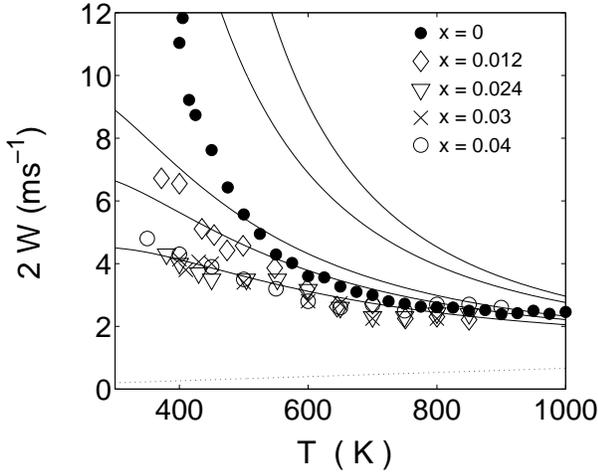}
\caption{Recalculated plane copper nuclear spin-lattice relaxation rate
$^{63}(1/T_1) = 2W$ with fixed $\zeta=1$ is shown by solid lines from up
till down with increasing doping. The dashed line shows the contribution to
$^{63}(1/T_1)$ from spin diffusion for $x=0.04$ with $\zeta=1$. The notation
is the same as in Fig.~\ref{Rel_63Cu}.}
\label{Rel_63Cu_zeta1}
\end{figure}

One may wonder about the results of calculations if we avoid fitting
$^{63}(1/T_1)$. Figure \ref{Rel_63Cu_zeta1} shows the results of
calculations for $^{63}(1/T_1)$ with fixed decoupling parameter $\zeta = 1$.
It is seen, that the theory is able to reproduce the main features of doping
and temperature dependence of $^{63}(1/T_1)$, however, in poor agreement
with respect to the numerical values.

It is interesting to compare the results of the calculations for carrier
free AF system with other theories. After the seminal work,\cite{CHN}
Chakravarty and Orbach\cite{ChakravartyOrbach} predicted the decrease of
$^{63}(1/T_1)$ with increasing temperature at low temperatures as
$^{63}(1/T_1) \sim T^{3/2} \exp(2\pi \rho_S/k_B T)$. After passing through a
wide minimum at $T \sim J/2$ the calculated $^{63}(1/T_1)$ {\it increases}
with $T$ at high temperatures where the system is recognized to be in the
quantum critical (QC) region. On the other hand, using the 1/N expansion
method on the N-component nonlinear sigma model an apparent formula was
obtained for $^{63}(1/T_1)$ in the QC region.\cite{ChubSachYe} However, the
behavior of $^{63}(1/T_1)$ was found to be nearly independent on
temperature: $^{63}(1/T_1) \sim (T/J)^{0.028}$. The calculations with
thermally excited skyrmions by Belov and Kochelaev\cite{Kochelaev_Belov_1T1}
showed that $^{63}(1/T_1)$ has also the nonmonotonic temperature dependence
with the slight increase at $T > 0.67 J$. In the present theory the
calculated value of contribution to $^{63}(1/T_1)$ relaxation rate from spin
diffusion at $T = 1000$~K is 550 sec$^{-1}$ in perfect agreement with
300-600 sec$^{-1}$ as estimated by Sokol {\it et al.}\cite{Sokol_Spin_Diff}
using exact diagonalization technique. However, their estimate was based on
the presence of disorder in CuO$_2$ planes.

  %%%   Figure 11

  %%%   Figure 12

   %%%   Figure 13

    %%%   Figure 14

    %%%   Figure 15

\section{Conclusion}

The theory for relaxation function in the two-dimensional $t-J$ model in the
paramagnetic state is presented taking into account the hole subsystem as
well as both the electron and antiferromagnetic correlations. The
presentation obeys rotational symmetry of the spin correlation functions and
is valid for all wave vectors through the Brillouin zone. The fit of
effective correlation length $\xi_{\mathit eff}$ to experimental data is in
agreement with the microsegregation hypothesis, where the effect of doped
holes affects the value of correlation length at $T \rightarrow 0$ as,
$\xi_0 = 1/nx$, where $n$ is the average distance between the holes along
the domain walls. The best fit yields $n=1.3$ for samples with $x \lesssim
0.02$ and $n=2$ for samples with $x > 0.02$. The expression for fourth
frequency moment of relaxation shape function is derived within the $t-J$
model. The spin diffusion contribution to relaxation rates is evaluated and
is shown to play a significant role in carrier free and doped
antiferromagnet in agreement with exact diagonalization calculations. The
convergence of contribution from spin diffusion to spin-lattice relaxation
rates is preserved by linear response theory and hydrodynamics. At low
temperatures the nuclear spin-lattice relaxation rate, $^{63}(1/T_1)$, of
plane $^{63}$Cu has the main contribution from the AF wavevector
$(\pi,\pi)$, and the $^{17}(1/T_1)$, of plane $^{17}$O, has the
contributions from the wavevectors in the vicinity of $(\pi,\pi)$ and small
$q \sim 0$. It is shown that the theory is able to explain the main features
of experimental data on temperature and doping dependence of copper nuclear
spin-lattice relaxation rate in both carrier free La$_2$CuO$_4$ and doped
La$_{2-x}$Sr$_x$CuO$_4$ compounds.

\section{Acknowledgments}

It is a pleasure to thank Dr.~A.Yu.~Zavidonov for bringing my interest to
the present study, Professor M.V.~Eremin for discussions, and Professor
A.~Rigamonti for discussions and hospitality in Pavia University, where part
of this work was carried out.

This work was supported in part by INTAS 01-0654 and joint US CRDF - RF
Minobr Grant $\#$~Y1-P-07-19.

\appendix{APPENDIX A}

Various types of sums over lattice sites utilized for calculation of
frequency moments $\left< {\omega^2_{\bf k}} \right> $ and $\left<
{\omega^4_{\bf k}} \right> $ in the $t-J$ model. The $G_{ij}$ and $D_{ij}$
symbols refer to hopping $t$ and(or) exchange $J$ between the first
neighbors, whereas $M$ and $Q$ refer to spin-spin correlation functions $c$
and(or) transfer amplitudes $T$ at the appropriate positions. $\delta_{ij}$
is the delta function. The sums for $M$ and $Q$ were calculated for up to
$n=|j-i|$-th neighbors ($n$ = 2).

\[ \gamma_{\bf k} \equiv (\cos k_x a + \cos k_y a)/2. \]

\[
\frac{1}{N} \sum_{ijlmr}
\delta_{im} \delta_{mr} \delta_{lr} D_{rj}^2 G_{rj} M_{rj}
e^{i {\mathbf{k}} \cdot \left( {\mathbf{R}}_{m} -{\mathbf{R}}_{r} \right) }
= 4 M_1 D^2 G ,
\]

\[
\frac{1}{N} \sum_{ijlmr} \delta_{mr} \delta_{ir} G_{lj} D_{rj}^2 M_{lj}
e^{i {\mathbf{k}} \cdot \left( {\mathbf{R}}_{m} -{\mathbf{R}}_{r} \right) }
= 16 M_1 D^2 G,
\]

\[
\frac{1}{N} \sum_{ijlmr} \delta_{ir} G_{lm} G_{mr} G_{rj} M_{lj} M_{mr}
e^{i {\mathbf{k}} \cdot \left( {\mathbf{R}}_{m} -{\mathbf{R}}_{r} \right) }
= 36 M_1^2 G^3 \gamma_{\bf k} ,
\]

\[
\frac{1}{N} \sum_{ijlmr} \delta_{mr} G_{il} G_{lr} G_{rj} M_{lr} M_{ij}
e^{i {\mathbf{k}} \cdot \left( {\mathbf{R}}_{m} -{\mathbf{R}}_{r} \right) }
= 36 M_1^2 G^3 ,
\]

\[
\frac{1}{N} \sum_{ijlmr} \delta_{mj} G_{il} G_{lm} G_{mr} M_{lm} M_{ir}
e^{i {\mathbf{k}} \cdot \left( {\mathbf{R}}_{m} -{\mathbf{R}}_{r} \right) }
= 36 M_1^2 G^3 \gamma_{\bf k} ,
\]

\[
\frac{1}{N} \sum_{ijlmr} \delta_{mj} \delta_{lr} \delta_{im} G_{mr}^3 M_{mr}
e^{i {\mathbf{k}} \cdot \left( {\mathbf{R}}_{m} -{\mathbf{R}}_{r} \right) }
= 4 M_1 G^3 \gamma_{\bf k} ,
\]

\[
\frac{1}{N} \sum_{ijlmr} \delta_{ir} \delta_{lm} D_{mr}^2 G_{rj} M_{rj}
e^{i {\mathbf{k}} \cdot \left( {\mathbf{R}}_{m} -{\mathbf{R}}_{r} \right) }
= 16 M_1 D^2 G \gamma_{\bf k} ,
\]

\[
\frac{1}{N} \sum_{ijlmr} \delta_{im} \delta_{lm} D_{mj}^2 G_{rj} M_{rj}
e^{i {\mathbf{k}} \cdot \left( {\mathbf{R}}_{m} -{\mathbf{R}}_{r} \right) }
= 16 M_1 D^2 G \gamma_{\bf k}^2 ,
\]

\[
\frac{1}{N} \sum_{ijlmr} \delta_{ir} \delta_{mr} D_{lr}^2 G_{rj} M_{rj}
e^{i {\mathbf{k}} \cdot \left( {\mathbf{R}}_{m} -{\mathbf{R}}_{r} \right) }
= 16 M_1 D^2 G ,
\]

\[
\frac{1}{N} \sum_{ijlmr} \delta_{mj} \delta_{il} D_{lm}^2 G_{mr} M_{mr}
e^{i {\mathbf{k}} \cdot \left( {\mathbf{R}}_{m} -{\mathbf{R}}_{r} \right) }
= 16 M_1 D^2 G \gamma_{\bf k} ,
\]

\[
\frac{1}{N} \sum_{ijlmr} \delta_{ir} G_{lm} G_{mr} G_{rj} M_{rj} M_{lm}
e^{i {\mathbf{k}} \cdot \left( {\mathbf{R}}_{m} -{\mathbf{R}}_{r} \right) }
= 64 M_1^2 G^3 \gamma_{\bf k} ,
\]

\[
\frac{1}{N} \sum_{ijlmr} \delta_{lr} G_{ir} G_{im} G_{rj} M_{im} M_{rj}
e^{i {\mathbf{k}} \cdot \left( {\mathbf{R}}_{m} -{\mathbf{R}}_{r} \right) }
= 64 M_1^2 G^3 \gamma_{\bf k}^2 ,
\]

\[
\frac{1}{N} \sum_{ijlmr} \delta_{lj} G_{ij} G_{im} G_{rj} M_{im} M_{rj}
e^{i {\mathbf{k}} \cdot \left( {\mathbf{R}}_{m} -{\mathbf{R}}_{r} \right) }
= 64 M_1^2 G^3 \gamma_{\bf k}^3 ,
\]

\[
\frac{1}{N} \sum_{ijlmr} \delta_{lj} G_{im} G_{mj} G_{rj} M_{im} M_{rj}
e^{i {\mathbf{k}} \cdot \left( {\mathbf{R}}_{m} -{\mathbf{R}}_{r} \right) }
= 64 M_1^2 G^3 \gamma_{\bf k}^2 ,
\]

\[
\frac{1}{N} \sum_{ijlmr} \delta_{lj} \delta_{ir} G_{mr} D_{rj}^2 Q_{mj}
e^{i {\mathbf{k}} \cdot \left( {\mathbf{R}}_{m} -{\mathbf{R}}_{r} \right) }
= 4 D^2 G \left( 3 Q_2 + Q_0 \right) \gamma_{\bf k},
\]

\[
\frac{1}{N} \sum_{ijlmr} \delta_{mr} \delta_{ij} G_{lj} D_{rj}^2 Q_{lr}
e^{i {\mathbf{k}} \cdot \left( {\mathbf{R}}_{m} -{\mathbf{R}}_{r} \right) }
= 4 D^2 G \left( 3 Q_2 + Q_0 \right),
\]

\[
\frac{1}{N} \sum_{ijlmr} \delta_{mr} G_{lr} G_{ir} G_{rj} M_{ij} M_{lr}
e^{i {\mathbf{k}} \cdot \left( {\mathbf{R}}_{m} -{\mathbf{R}}_{r} \right) }
= 16M_1 G^3 \left( 3M_2 + M_0 \right),
\]

\[
\frac{1}{N} \sum_{ijlmr} \delta_{ir} G_{lm} G_{mr} G_{rj} M_{mj} M_{lr}
e^{i {\mathbf{k}} \cdot \left( {\mathbf{R}}_{m} -{\mathbf{R}}_{r} \right) }
= 4 G^3 \left( 3M_2 + M_0 \right)^2 \gamma_{\bf k} ,
\]

\[
\frac{1}{N} \sum_{ijlmr} \delta_{ij} G_{lm} G_{mj} G_{rj} M_{mr} M_{lj}
e^{i {\mathbf{k}} \cdot \left( {\mathbf{R}}_{m} -{\mathbf{R}}_{r} \right) }
= 4 G^3 \left( 3M_2 + M_0 \right)
\left( M_2 \left( 4 \gamma_{\bf k}^2-1 \right) + M_0 \right) ,
\]

\begin{eqnarray}
\frac{1}{N} \sum_{ijlmr} \delta_{ij} & D_{lj} & D_{lm} G_{rj} Q_{mj} M_{lr}
e^{i {\mathbf{k}} \cdot \left( {\mathbf{R}}_{m} -{\mathbf{R}}_{r} \right) }
= 64 Q_2 M_2 D^2 G \gamma_{\bf k}^3 \nonumber \\
 + & 4 & \left(Q_0 M_0 + 3Q_0 M_2 + 3M_0 Q_2-7M_2 Q_2 \right) G^3
 \gamma_{\bf k} ,
\nonumber
\end{eqnarray}

\newpage

 .

 \newpage

\[
\frac{1}{N} \sum_{ijlmr} \delta_{lj} \delta_{ir} D_{mr} G_{rj} D_{rj} Q_{mj}
e^{i {\mathbf{k}} \cdot \left( {\mathbf{R}}_{m} -{\mathbf{R}}_{r} \right) }
= 4 D^2 G (3Q_2 + Q_0) \gamma_{\bf k} ,
\]

\[
\frac{1}{N} \sum_{ijlmr} \delta_{lj} G_{ij} G_{mj} G_{rj} M_{im} M_{rj}
e^{i {\mathbf{k}} \cdot \left( {\mathbf{R}}_{m} -{\mathbf{R}}_{r} \right) }
= 16 M_1 \left( 3M_2 + M_0 \right) G^3 \gamma_{\bf k}^2 ,
\]

\[
\frac{1}{N} \sum_{ijlmr} \delta_{mr} G_{il} G_{lr} G_{rj} M_{ir} M_{lj}
e^{i {\mathbf{k}} \cdot \left( {\mathbf{R}}_{m} -{\mathbf{R}}_{r} \right) }
= 4 \left( 3M_2 + M_0 \right)^2 G^3 ,
\]

\[
\frac{1}{N} \sum_{ijlmr} \delta_{mj} G_{il} G_{lm} G_{mr} M_{im} M_{lr}
e^{i {\mathbf{k}} \cdot \left( {\mathbf{R}}_{m} -{\mathbf{R}}_{r} \right) }
= 4 \left( 3M_2 + M_0 \right)^2 G^3 \gamma_{\bf k} ,
\]

\[
\frac{1}{N} \sum_{ijlmr} \delta_{lr} G_{ir} G_{mr} G_{rj} M_{ir} M_{mj}
e^{i {\mathbf{k}} \cdot \left( {\mathbf{R}}_{m} -{\mathbf{R}}_{r} \right) }
= 16 M_1 \left( 3M_2 + M_0 \right) G^3 \gamma_{\bf k} ,
\]

\[
\frac{1}{N} \sum_{ijlmr} \delta_{lj} G_{ij} G_{im} G_{rj} M_{ij} M_{mr}
e^{i {\mathbf{k}} \cdot \left( {\mathbf{R}}_{m} -{\mathbf{R}}_{r} \right) }
= 36 M_1^2 G^3 \gamma_{\bf k} ,
\]

\[
\frac{1}{N} \sum_{ijlmr} \delta_{mj} G_{lm} G_{im} G_{mr} M_{ir} M_{lm}
e^{i {\mathbf{k}} \cdot \left( {\mathbf{R}}_{m} -{\mathbf{R}}_{r} \right) }
= 16M_1 G^3 \left( 3M_2 + M_0 \right) \gamma_{\bf k},
\]

\[
\frac{1}{N} \sum_{ijlmr} \delta_{lr} G_{ir} G_{im} G_{rj} M_{ir} M_{mj}
e^{i {\mathbf{k}} \cdot \left( {\mathbf{R}}_{m} -{\mathbf{R}}_{r} \right) }
= M_1^2 G^3 \left( 12 + 16 \gamma_{\bf k}^2 + 8 \cos k_x a \cos k_y a \right),
\]

\[
\frac{1}{N} \sum_{ijlmr} \delta_{ij} G_{lm} G_{mj} G_{rj} M_{lr} M_{mj}
e^{i {\mathbf{k}} \cdot \left( {\mathbf{R}}_{m} -{\mathbf{R}}_{r} \right) }
= M_1^2 G^3 \left( 12 + 16 \gamma_{\bf k}^2 + 8 \cos k_x a \cos k_y a \right),
\]

\[
\frac{1}{N} \sum_{ijlmr} \delta_{lj} G_{ij} G_{mj} G_{rj} M_{ij} M_{mr}
e^{i {\mathbf{k}} \cdot \left( {\mathbf{R}}_{m} -{\mathbf{R}}_{r} \right) }
= 16 M_1 G^3 \left( M_2 \left( 4 \gamma_{\bf k}^2-1 \right) + M_0 \right) ,
\]

\[
\frac{1}{N} \sum_{ijlmr} \delta_{lj} G_{ij} G_{im} G_{rj} M_{mj} M_{ir}
e^{i {\mathbf{k}} \cdot \left( {\mathbf{R}}_{m} -{\mathbf{R}}_{r} \right) }
= 64 M_2^2 G^3 \gamma_{\bf k}^3 +
  4 \left( M_0^2 + 6 M_0 M_2 - 7 M_2^2 \right) G^3 \gamma_{\bf k} ,
\]

\[
\frac{1}{N} \sum_{ijlmr} \delta_{mj} G_{lm} G_{im} G_{mr} M_{il} M_{mr}
e^{i {\mathbf{k}} \cdot \left( {\mathbf{R}}_{m} -{\mathbf{R}}_{r} \right) }
= 16 M_1 \left( 3M_2 + M_0 \right) G^3 \gamma_{\bf k} ,
\]

\[
\frac{1}{N} \sum_{ijlmr} \delta_{ir} G_{lr} G_{mr} G_{rj} M_{lm} M_{rj}
e^{i {\mathbf{k}} \cdot \left( {\mathbf{R}}_{m} -{\mathbf{R}}_{r} \right) }
= 16 M_1 \left( 3M_2 + M_0 \right) G^3 \gamma_{\bf k} ,
\]

\[
\frac{1}{N} \sum_{ijlmr} \delta_{lr} \delta_{ij} D_{mj} G_{rj} D_{rj} Q_{mr}
e^{i {\mathbf{k}} \cdot \left( {\mathbf{R}}_{m} -{\mathbf{R}}_{r} \right) }
= 4 D^2 G ( Q_2 (4\gamma_k^2-1)+Q_0) ,
\]

\newpage

 .

 \newpage

\appendix{APPENDIX B}

In this appendix the relations to compare the result for frequency moments
of $F({\bf k}, {\omega}) $ in carrier free 2DHAF are given.

For the second and fourth moments the result was\cite{LoveseyMeserve}
\begin{equation} \label{chinaW2}
\left< \omega^2_{\bf k} \right> _{{\rm 2DHAF}}=-\frac {8J(1-\gamma_{\bf k})}
 {\chi({\bf k})N} \sum_{{\bf q}} \gamma_{\bf q} S({\bf q}),
\end{equation}
and
\begin{equation} \label{chinaW4}
 \left< \omega^4_{\bf k} \right> _{{\rm 2DHAF}} =
- \frac {128 J^3 } {\chi({\bf k}) N^2} \sum_{{\bf pq}} S({\bf p})S({\bf q})
F({\bf k},{\bf p},{\bf q}),
\end{equation}
where
\begin{eqnarray} \label{chinaFkpq}
\cr & \cr & F({\bf k},{\bf p},{\bf q}) \simeq \gamma_{\bf k}
\left(-2\gamma_{\bf p}\gamma_{\bf q} - 10 \gamma_{\bf p} \gamma_{\bf q}^2
-4\gamma_{\bf p}\gamma_{\bf q}^3 +\gamma_{\bf p}^2 \gamma_{\bf q}^2 \right)
\cr & \cr & + \gamma_{\bf k}^2 \left( 4\gamma_{\bf p} \gamma_{\bf q}
+ 3 \gamma_{\bf p} \gamma_{\bf q}^2 \right)
- 2 \gamma_{\bf k}^3\gamma_{\bf p}\gamma_{\bf q}
+ 5\gamma_{\bf p} \gamma_{\bf q}^2
\cr & \cr & - \gamma_{\bf p}^2\gamma_{\bf q}^2 + \gamma_{\bf p}
\gamma_{\bf q}^3 +5\gamma_{\bf p} \gamma_{\bf q} \gamma_{{\bf k}-{\bf q}}^2
+ \gamma_{{\bf k}+{\bf p}}^2 \gamma_{\bf q}^2
\cr & \cr & + 2 \gamma_{\bf p} \gamma_{{\bf k}-{\bf q}}^2
- 2\gamma_{\bf p}\gamma_{{\bf k}-{\bf q}}^3 - \gamma_{{\bf k}+{\bf p}}
\gamma_{{\bf k}-{\bf q}} \gamma_{{\bf k}+{\bf p}-{\bf q}} .
\end{eqnarray}
Using the relation between correlation functions in momentum and site
representation:

\begin{eqnarray} \label{Sk_c0c1c2}
S({\bf k}) & \equiv & \langle S_{\bf k}^z S_{-{\bf k}}^z \rangle =
%%% \frac {1}{N}
\sum_{|i-j| \leq 2} \langle S_i^z S_j^z \rangle
e^{i {\mathbf{k}} \cdot \left( {\mathbf{R}}_i-{\mathbf{R}}_j \right) } =
\cr & = & c_0 + 4 c_1 \gamma_{\bf k} + 4 c_2 (4 \gamma_{\bf k}^2
- \cos k_x a \cos k_y a - 1 ) ,
\end{eqnarray}
performing the summation over ${\bf k}$ and ${\bf q}$ in (\ref{chinaW4}) and
using the necessary types of integrals (sums) over the Brillouin zone as
given below, we obtain the same expression, as shown in (\ref{w4decoupled}),
with $t_{\mathit eff}=0$ and settling to unity the decoupling parameter
$\zeta = 1$. The integrals (sums) were calculated analytically and checked
numerically.

\[ \gamma_{\bf q} \equiv (\cos q_x a + \cos q_y a)/2. \]
\[
\sum_{{\bf p}}\gamma_{\bf p}^4=\frac {9}{64},
\mbox{\hspace{5mm}}
\sum_{{\bf p}}\gamma_{\bf p}^2=\frac14,
\mbox{\hspace{5mm}}
\sum_{{\bf q}}\gamma_{{\bf k}-{\bf q}}^2=\frac 14,
\]

\[
\sum_{{\bf p},{\bf q}} \cos p_x a \cos p_y a \gamma_{{\bf k} +{\bf
p}}\gamma_{{\bf k}-{\bf q}} \gamma_{{\bf k}+{\bf p}-{\bf q}}
=\frac {1}{32} \gamma_{\bf k},
\]

\[
\sum_{{\bf p},{\bf q}} \gamma_{\bf p}^2 \gamma_{\bf q}^2 \gamma_{{\bf k}+
{\bf p}} \gamma_{{\bf k}-{\bf q}} \gamma_{{\bf k}+{\bf p}- {\bf q}}=\frac
{1}{256} \left( \gamma_{\bf k}^3 +\frac 54 \gamma_{\bf k} \cos k_x a \cos
k_y a +\frac {45}{16} \gamma_{\bf k} \right),
\]

\[
\sum_{{\bf p},{\bf q}}
\cos p_x a \cos p_y a \gamma_{\bf q}^2 \gamma_{{\bf k}+
{\bf p}} \gamma_{{\bf k}-{\bf q}}\gamma_{{\bf k}+{\bf p}-{\bf q}}
= \frac {3 \gamma_{\bf k}}{512} \left( 2 \cos k_x a \cos k_y a + 1 \right),
\]

\[
\sum_{{\bf p},{\bf q}} \cos p_x a \cos p_y a \cos q_x a \cos q_y a
\gamma_{{\bf k}+{\bf p}} \gamma_{{\bf k}-{\bf q}}
\gamma_{{\bf k}+{\bf p}-{\bf q}}
= \frac {1}{64}  \cos k_x a \cos k_y a \gamma_{\bf k},
\]

\[
 \sum_{{\bf p}}\gamma_{\bf p}=0, \mbox{\hspace{5mm}} \sum_{{\bf q}}
 \gamma_{\bf q} \gamma_{{\bf k}-{\bf q}}^2=0, \mbox{\hspace{5mm}} \sum_{{\bf
 p}}\gamma_{\bf p}^3=0,
\]

\[
\sum_{{\bf p}} \gamma_{\bf p} \cos p_x a \cos p_y a = 0,
\mbox{\hspace{7mm}}
\sum_{{\bf q}} \gamma_{\bf q}^3 \gamma_{{\bf k}-{\bf q}}^2=0,
\]

\[
\sum_{\bf q} \gamma_{\bf q}^2 \gamma_{{\bf k}-{\bf q}}^2 = \frac {1}{16}
\left( \frac 34 +\frac 12 \cos k_x a \cos k_y a + \gamma_{\bf k}^2 \right),
\]

\[
\sum_{{\bf q}} \gamma_{\bf q} \gamma_{{\bf k}-{\bf q}}^3
= \frac {9}{64} \gamma_{\bf k}, \mbox{\hspace{2mm}}
\sum_{{\bf q}} \gamma_{\bf q}^2 \cos q_x a \cos q_y a = \frac 18,
\]

\[
\sum_{{\bf q}} \gamma_{{\bf k}-{\bf q}}^2 \cos q_x a \cos q_y a
=\frac 18 \cos k_x a \cos k_y a,
\]

\[
\sum_{{\bf p},{\bf q}} \gamma_{{\bf k}+{\bf p}}
\gamma_{{\bf k}-{\bf q}}\gamma_{{\bf k}+{\bf p}-{\bf q}}
=\frac {1}{16} \gamma_{\bf k},
\]

\[
\sum_{{\bf p},{\bf q}} \gamma_{\bf p}^2 \gamma_{{\bf k}+ {\bf
p}}\gamma_{{\bf k}-{\bf q}} \gamma_{{\bf k}+{\bf p}-{\bf q}}
=\frac {9}{256} \gamma_{\bf k},
\]

\end{document}